\documentclass[aps,prl,reprint,groupedaddress,nobibnotes]{revtex4-2}
\usepackage{graphicx}
\usepackage{upgreek}
\usepackage{amsmath}
\usepackage{hyperref}
\usepackage{xr}
\usepackage[version=4]{mhchem}
\usepackage{cleveref}
\begin{document}


\title{Electrochemical Shot Noise of a Redox Monolayer }


\author{Simon Grall, Shuo Li, Laurent Jalabert, Soo-Hyeon Kim}
\affiliation{IIS, LIMMS/CNRS-IIS IRL2820, The Univ. of Tokyo; 4-6-1 Komaba, Meguro-ku Tokyo, 153-8505, Japan}
\author{Arnaud Chovin, Christophe Demaille} 
\email[Christophe Demaille: ]{christophe.demaille@univ-paris-diderot.fr}

\affiliation{Université Paris Cité, CNRS, Laboratoire d'Electrochimie Moléculaire, F-75013 Paris, France}
\author{Nicolas Cl\'ement} 
\email[Nicolas Cl\'ement: ]{nclement@iis.u-tokyo.ac.jp}
\affiliation{IIS, LIMMS/CNRS-IIS IRL2820, The Univ. of Tokyo; 4-6-1 Komaba, Meguro-ku Tokyo, 153-8505, Japan}



\date{\today}

\begin{abstract}
Redox monolayers are the base for a wide variety of devices including high-frequency molecular diodes or biomolecular sensors. We introduce a formalism to describe the electrochemical shot noise of such monolayer, confirmed experimentally at room temperature in liquid. The proposed method, carried out at equilibrium, avoids parasitic capacitance, increases the sensitivity and allows to obtain quantitative information such as the electronic coupling (or standard electron transfer rates), its dispersion and the number of molecules. Unlike in solid-state physics, the homogeneity in energy levels and transfer rates in the monolayer yields a Lorentzian spectrum. This first step for shot noise studies in molecular electrochemical systems opens perspectives for quantum transport studies in liquid environment at room temperature as well as highly sensitive measurements for bioelectrochemical sensors.

\end{abstract}


\maketitle

Self-assembled monolayers (SAM) composed of nanometric-long redox molecules are building blocks for molecular electronics and electrochemistry. They can behave as molecular diodes operating at ultra-high-frequency (potentially as rectenna in the visible spectrum) \cite{trasobares_17_2016,reynaud_rectifying_2020}, with on-off ratio breaking the Landauer limit \cite{clement_breaking_2017, chen_molecular_2017}, and show interesting features such as signatures of collective quantum interference effects at room temperature \cite{trasobares_estimation_2017,li_large_2022,gehring_single-molecule_2019}. In addition, their operation in liquid offers a direct link between quantum transport and electrochemistry \cite{bevan_exploring_2016,nitzan_electron_2003,marcus_theory_1956} that provides unique opportunities. For example, the nanoscale measurements of electrochemical signals remains extremely challenging while key to the development of nanobiosensors \cite{li_redox-labelled_2022}. Several approaches have been explored to tackle the challenge, using redox cycling \cite{fan_single_1996}, high frequency measurements \cite{grall_attoampere_2021} and fluorescence \cite{huang_high-throughput_2015}. The underlying challenges rise from the presence of parasitic capacitances and from the fact that under typical measurements conditions, the current scales with the sensor area, leading to difficulties in retrieving the signal with micro- and nanoscale electrodes. Simultaneously, these systems offer unique properties as quantum devices. Probably the most intriguing aspect for the solid-state physics community is the potential for millions of single-energy level quantum dots simultaneously operating at room temperature, with extremely small dispersion, tunable electronic coupling \cite{chidsey_free_1991} and Frank Condon effect \cite{demaille_electrochemical_2022}.\\
We propose here to exploit and formalize the shot noise induced by reversible single electron transfers of electroactive molecules attached to an electrode as a new, very sensitive electrochemical technique and as a way to characterize the homogeneity in the electronic properties of these assembled molecular quantum dots. shot noise has been extensively studied in solid-state physics \cite{blanter_shot_2000} and more recently in molecular electronics \cite{djukic_shot_2006,lumbroso_electronic_2018}, but not in electrochemistry, except for the shot noise due to a variation of the number of molecules in a nanogap \cite{zevenbergen_fast_2009,mathwig_electrical_2012,katelhon_noise_2013,singh_stochastic_2016}. Such measurements are usually challenging because of the ubiquitous $1/f$ noise (e.g. in solid-state physics \cite{karnatak_1_2017}, quantum transport \cite{paladino_1_2014}, molecular electronics \cite{clement_1_2007} or in liquid \cite{fragasso_comparing_2020}) which is typically circumvented by low-temperature measurements and by measurements at higher relative frequencies. \\
The $1/f$ noise is here not dominant thanks to the well-defined energy level and electron transfer rates of the redox molecules of the monolayer, allowing to study its low-frequency shot noise arising from the sum of single-electrons trapping/detrapping events to each molecule with a narrow distribution in time constants.  A simple and straight-forward equation of the shot noise is proposed, giving direct access to the distribution of the charge transfer rates and the number of charge carriers. This approach provides clearly readable signals even when faradaic currents become unmeasurable, avoids the parasitic capacitance issue and allows for measurements without extra excitation other than the thermal noise.\\


\begin{figure*} 
\includegraphics[trim={2cm 1cm 2cm 1cm},clip,width=\textwidth]{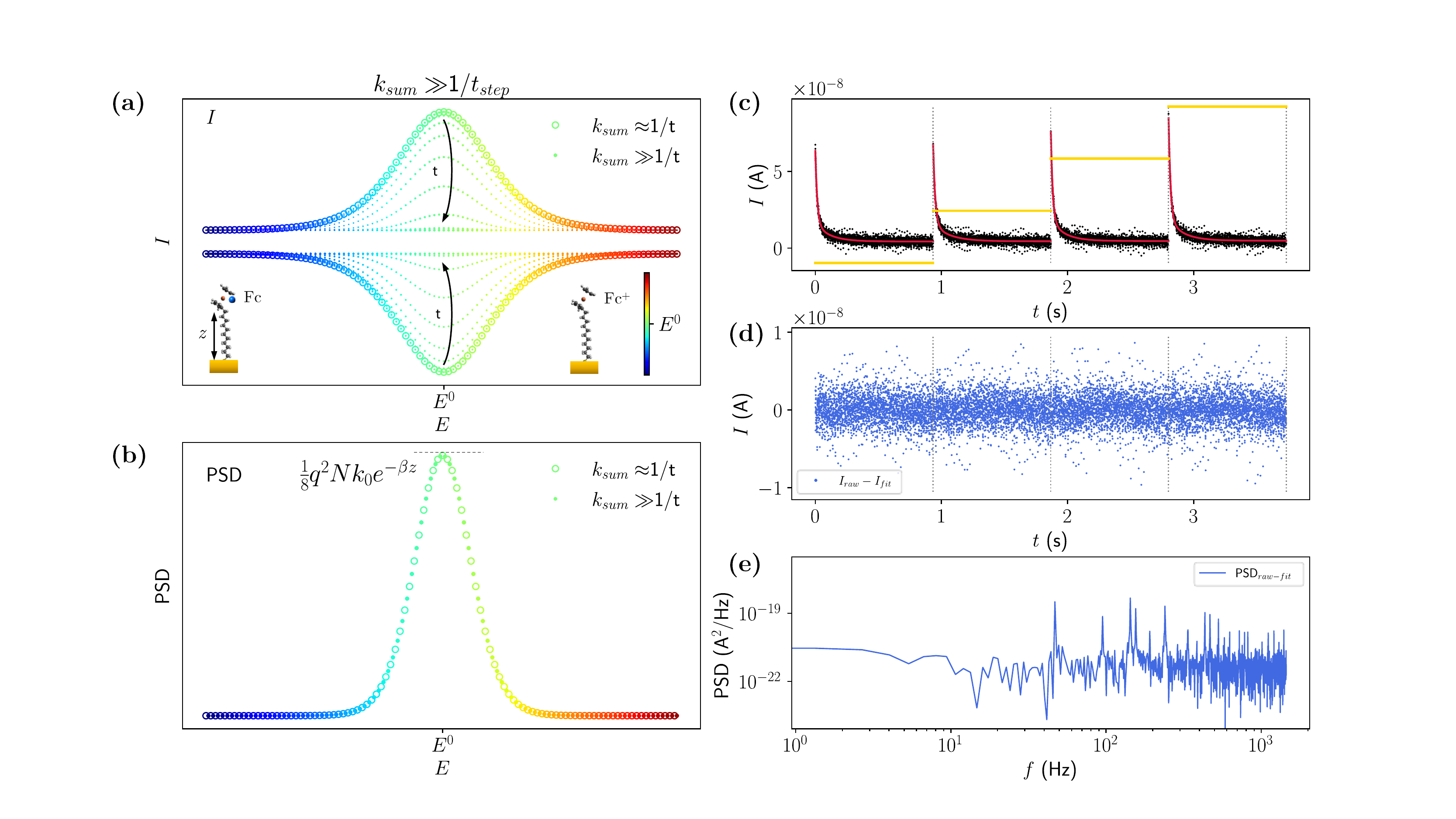}%
\caption{Illustration of current and noise versus time and voltage, considering a slow scan rate compared to the electron transfer rates ($k_{sum}\gg 1/t_{step}$). (a) $I$ vs $E$, with the evolution of $I$ as the time after voltage step is increased. (b) Voltnoisograms (PSD vs $E$) taken at low frequency (Eq. \ref{S_f0}) corresponding to the same conditions as in (a). (c) Sampled staircase voltammetry example, with the raw current data (black dots), a double exponential decay fit of the current (red) and the voltage steps (yellow). In this example, each voltage step is of 2.25 mV, starting at 195 mV. (d) Raw currents subtracted with exponential fits (blue). (e) PSD spectrum of one current timetrace in (d). \label{figure1}}
\end{figure*}

Electroactive redox molecules can be seen as single-electron quantum dots with extremely small energy dispersion, even in liquid and at ambient temperature \cite{trasobares_estimation_2017}. The equilibrium reaction of an ideally reversible redox couple M$^+$/M attached to a metallic electrode and held at a distance $z$ from the electrode (insets Figure \ref{figure1} (a)) can be written as:

\begin{center}
\ce{M <=>[k_{ox}][k_{red}] M^+ + e^-}
\end{center}

Using the Marcus-Hush formalism to describe the electron transfer rates gives \cite{smalley_kinetics_1995}:

\begin{equation}
 k^{MH}_{ox,red} = \frac{\rho H^2}{\hbar} \sqrt{\frac{\pi}{k_BT\lambda}} \int\limits_{-\infty}^{+\infty}\frac{1}{1+e^{\frac{x}{k_BT}}}e^{-\frac{(x-\lambda \pm \eta)^2}{4\lambda k_BT}}dx
\label{k_marcus}
\end{equation}
\noindent
with $k_{ox}$ the oxidation rate, $k_{red}$ the reduction rate, $\rho$ the density of state in the metallic electrode, $H^2$ the electronic coupling, $\hbar$ the reduced Planck constant, $\lambda$ the reorganization energy (Frank Condon effect due to water molecules reorganizing after charging the redox molecule), $T$ the temperature, $k_B$ the Boltzmann constant and $\eta=q(E-E^0)$ with $E$ the potential at the electrode, $E^0$ the standard potential of the molecule and $q$ the elementary charge. Eq.\ref{k_marcus} is analogue to the Landauer formalism in solid-state physics \cite{bevan_exploring_2016}. The specificity of the redox molecules is their energy level broadening due to the large reorganization energy. Eq.\ref{k_marcus} can be simplified to Eq. \ref{kox} \ref{kred} (Buttler-Volmer model) when $|\eta|<<\lambda$ which is often the case. It will be used here initially for its simplicity.

\begin{align}
 k^{BV}_{ox} = & k_0 e^{-\beta z} e^{\alpha \frac{\eta}{k_B T}}\label{kox}\\
 k^{BV}_{red} = & k_0 e^{-\beta z} e^{-(1-\alpha) \frac{\eta}{k_B T}}\label{kred}
\end{align}
\noindent
with $\beta$ the tunneling decay coefficient (1 Å$^{-1}$), $\alpha$ the charge transfer coefficient and $k_0$ the standard electron transfer rate at a distance $z$ = 0 (in s$^{-1}$). The exponential decay part is formally contained in the electronic coupling term $H^2$ in Eq. \ref{k_marcus} but is usually extracted for convenience to be included in the Butler-Volmer model \cite{smalley_kinetics_1995}.

Sampled current staircase voltammetry (SCV) is the electrochemical technique used to interrogate the surface-attached redox species \cite{heering_using_1999}, analogue to the charge pumping technique in semiconductors \cite{fujiwara_nanoampere_2008}. The electrode potential is raised in small steps of height $E_{step}$, and the current is recorded as a function of time, up to a time $t_{step}$, corresponding to the steps duration (Figure \ref{figure1} (c)). 

%
%
%
%

%
%
%

The current $I$, in the case of slow scan rate and long sampling time (i.e. $k_{sum}=k^{BV}_{ox}+k^{BV}_{red}\gg 1/t$), $I$ can be expressed as (details in SM):
\begin{equation} \label{I_simple}
I =  \frac{N q\nu}{4k_BT}\frac{1}{\cosh^2(\frac{\eta}{2k_BT})}	
\end{equation}
\noindent
with $N$ the total number of molecules and $\nu=E_{step}/t_{step}$ the voltage scan rate.
Note that such current represents the transition of the charges at a certain scan rate, and not an equilibrium value of the current at a given potential. Figure \ref{figure1} (a) shows $I$ versus applied voltage $E$ at a given scan rate and at different times $t$ after the voltage step, exhibiting a quick decrease of amplitude.\\

One way to consider the noise of the current versus time (Figure \ref{figure1} (b)) is to look at its power spectrum density (PSD, noted $S$ in equations). The PSD (Figure \ref{figure1} (b) and (e)) can be seen as a description of how the variance of the measured signal is spread in the frequency domain. The system under study has two-states related to the oxidized/reduced states of the molecules, here attached to a single electrode. Each molecule is expected to lead to the so-called Random Telegraph Signal (RTS) which is a shot noise due to individual transfer of electrons in and out of the single-electron boxes. To avoid confusion, in such a single-electrode system, the shot noise is not expected to be compared to  $2qI$ because at equilibrium, where both oxidation and reduction reactions compensate each other, $I=0$ while $S>0$ (discussion in SM) \cite{sarpeshkar_white_1993}. In general, RTS is typically associated with $1/f$ noise due to the wide range of energy levels and electron transfer rates\cite{machlup_noise_1954,clement_one-by-one_2010}. However, an ensemble of reversible redox couples, like those found in a redox SAM in liquid, can be thought of as an ensemble of quantum dots with very similar energy levels because the molecules that make up the SAM have strictly identical atomic structures and may differ only in their orientation relative to the surface\cite{trasobares_estimation_2017}. Assuming first that all $N$ molecules have identical energy levels $E^0$ and charge transfer rates $k_{ox}$/$k_{red}$ for oxidation/reduction, respectively, the PSD can be expressed as \cite{machlup_noise_1954}:

\begin{equation} \label{S_deltaI}
S(f,\eta,N) = 4N\Delta I^2  \frac{k_{ox}k_{red}}{k_{ox}+k_{red}} \frac{1}{(k_{ox}+k_{red})^2+(2\pi f)^2}	
\end{equation}

\noindent
with $f$ the frequency and $\Delta I$ the current corresponding to the oxidation (or reduction) of one molecule. If we consider $\Delta I$ as the transfer of one electron of charge $q$ per the average time taken for transferring one electron (i.e., $\Delta I=\frac{q}{\frac{1}{k_{ox}} +\frac{1}{k_{red}}}$), $S$ can be rewritten as:

\begin{equation} \label{S_full}
S(f,\eta,N) = 4N q^2 \frac{(k_{ox}k_{red})^3}{(k_{ox}+k_{red})^3} \frac{1}{(k_{ox}+k_{red})^2+(2\pi f)^2}	
\end{equation}
\noindent
which becomes at low frequency (assuming $\alpha=0.5$):

\begin{align}
			\lim_{f \rightarrow 0} S(\eta,N) 			&=4N q^2 \frac{(k_{ox}k_{red})^3}{(k_{ox}+k_{red})^5}\label{S_f0_k}\\
																						&=\frac{1}{8}Nq^2\frac{k_0e^{-\beta z}}{\cosh^5(\frac{\eta}{2k_bT})}\label{S_f0}
\end{align}
This equation expresses the dependence of the low frequency electrochemical shot noise of the redox SAM versus the electrode potential. The corresponding curve is plotted in Figure \ref{figure1} (b). Similarly to current SCV signals, it presents a peak at $E^0$, but narrower than that of the SCV peak, with a full width at half-maximum (FWHM):

\begin{equation}
E^S_{FWHM}=4\operatorname{acosh}(\sqrt[5]{2})\frac{k_BT}{q}\approx 56 \text{~mV}
\label{FWHM_S}
\end{equation}

Note that unlike the current, $S$ does not depend on $\nu$ as the PSD is considered for a system at equilibrium. $S$ is also independent of the potential scan direction. Interestingly, the limiting cases of $\eta \rightarrow 0$ and $f\rightarrow 0$ give access to the electron transfer rate $k_0$ and the total number of molecules $N$.

\begin{equation} \label{S_V0}
\lim_{\eta \rightarrow 0} S(f,N) = \frac{1}{2}N q^2 k_0 e^{-\beta z} \frac{1}{4+(\frac{2\pi f}{k_0 e^{-\beta z}})^2}	
\end{equation}

\noindent
with the corner frequency $f_c = \frac{1}{2\pi}k_0 e^{-\beta z}	$:

\begin{equation} \label{S_V0_f0}
\lim_{\eta \rightarrow 0, f\rightarrow 0} S(N) = \frac{1}{8}N q^2 k_0 e^{-\beta z} 
\end{equation}

The main result of the present work is Eq. \ref{S_V0_f0}, linking directly and simply $k_0$ and $N$ to the noise measured at low frequency for $E=E^0$. Provided the corner frequency of the PSD $f_c$ can be measured (Figure \ref{SM_FFT} (b)), the individual values of $k_0$ and $N$ are obtained from Eq. \ref{S_V0} and Eq. \ref{S_V0_f0}. Alternatively, if $N$ is known independently, $k_0$ can be straightforwardly derived from $S$ at $E=E^0$ (Eq. \ref{S_V0_f0}).\\

\begin{figure} 
\includegraphics[trim={6cm 1.4cm 4.5cm 1.4cm},clip,width=0.5\textwidth]{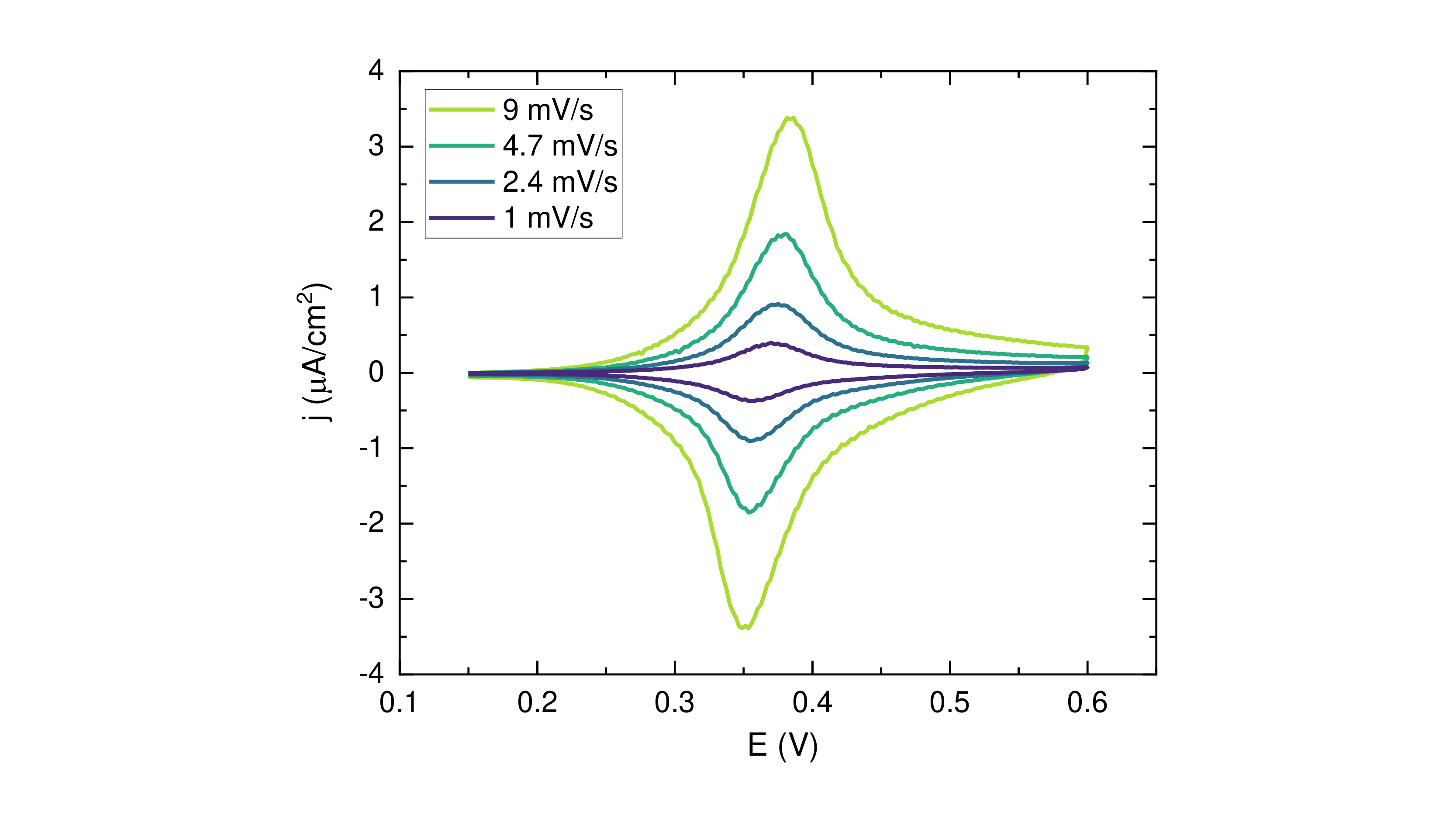}%
\caption{Example of current CVs obtained at different $\nu$ (electrode area $\approx 45$ mm$^2$, $E$ vs Ag/AgCl (3 M NaCl), electrolyte: [NaClO$_4$]=0.5 M). \label{CV_current}}
\end{figure}

To demonstrate the validity of the previous analysis, an experiment is set using ferrocene undecanethiol Fc(CH$_2$)$_{11}$SH self-assembled on a gold microelectrode. A two-electrode electrochemical cell setup is used in a Faraday cage, using a [NaClO$_4$]=0.5 M aqueous electrolyte and a Ag/AgCl electrode (3 M NaCl) acting as both reference and counter electrode. Details about the sample preparation and the measurement setup can be found in Supplementary Materials (Figure \ref{map_electrodes} and \ref{SM_measurement_setup}). The system is interrogated using staircase voltammetry (Figure \ref{figure1} (c)), which is equivalent to  linear cyclic voltammetry (CV) at slow scan rates \cite{christie_theory_1965}. Our motivation is to offer a comparison of the well-known technique of cyclic voltammetry with the results obtained looking at the shot noise of the system.


The Figure \ref{CV_current} shows an example of current CVs at different (low) scan rates $\nu$. The signal is centered around a potential value of $E^0=0.35 \pm 0.02$ V vs Ag/AgCl, which corresponds to the expected standard potential for such surface-attached Fc molecules \cite{tian_modulated_2013,nerngchamnong_nonideal_2015,gupta_role_2021}. The density is estimated here at 4.2$\times 10^{-10}$ mol/cm$^2$, close to the values reported in the literature for packed SAMs ($4.4 \sim 4.9\times 10^{-10}$ mol/cm$^2$) \cite{nijhuis_molecular_2009,trasobares_17_2016}. The peak current of the CV exhibit the usual behavior for a surface-confined reversible couple, with a linear dependency of the current versus $\nu$ (example data Figure \ref{SM_sweeprates}).





%
%
%

\begin{figure}
\includegraphics[trim={11.5cm 2cm 11.5cm 1cm},clip,width=0.45\textwidth]{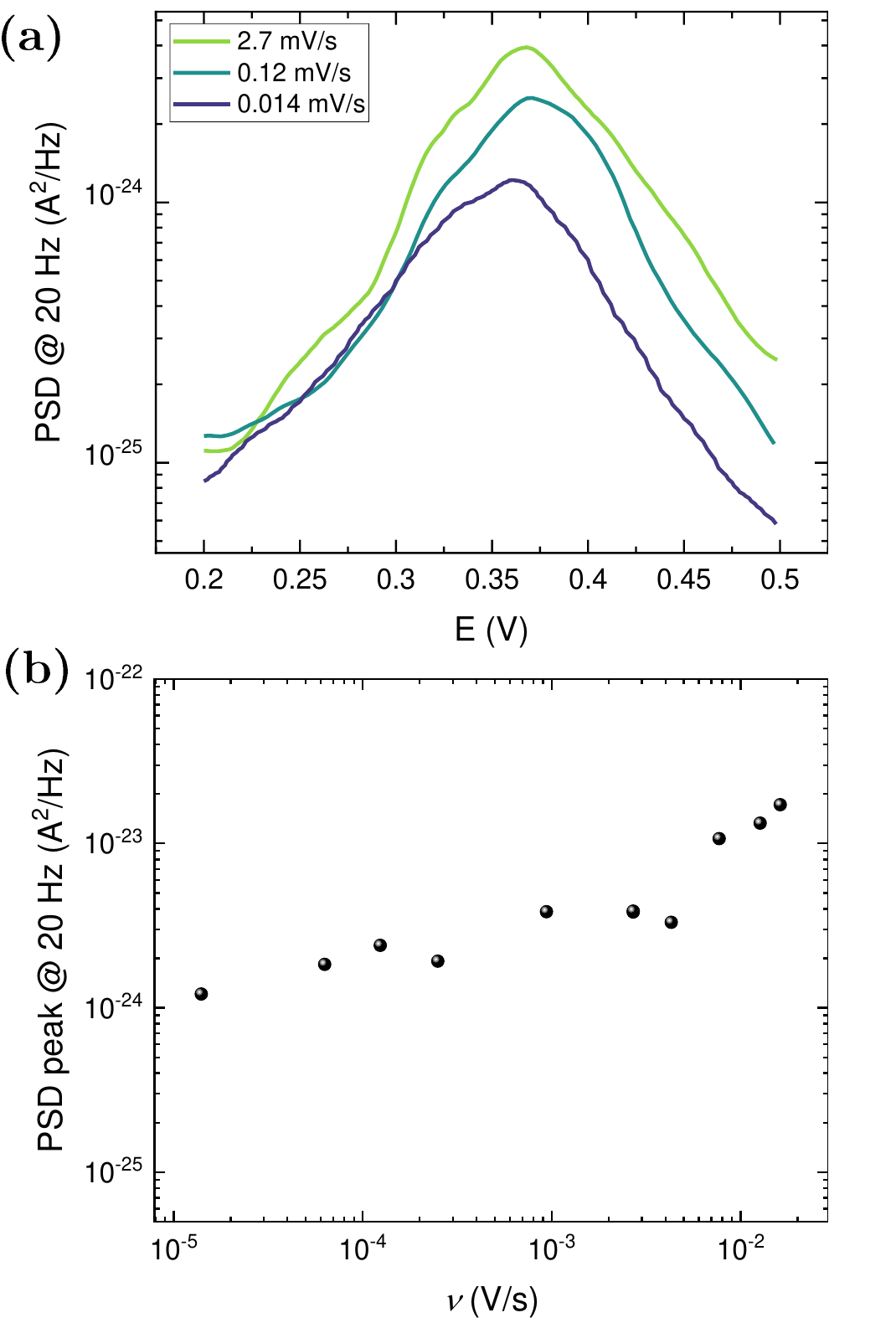}
\caption{(a) PSD of the current versus $E$ obtained at different $\nu$, at $f \approx$ 20 Hz. (b) PSD at $f \approx$ 20 Hz and $E=$ 0.35 V versus $\nu$.\label{CV_PSD}}
\end{figure}
%
PSD signals were measured at several scan rates (see details in SM), their magnitude at 20 Hz versus $E$ (called ``voltnoisogram'' for concision) shown in Figure \ref{CV_PSD} (a), (full set  Figure \ref{SM_sweeprates_PSD}). Similar data without the Fc molecules can be found in Figure \ref{SM_C11}. The PSD voltnoisograms behave as expected with a peak-shaped curve centered around $E^0\approx 0.35$ V, close to the standard potential of Fc. As predicted from Eq. \ref{S_f0}, the peak value of the PSD voltnoisograms (Figure \ref{CV_PSD} (b)) remains quasi-constant for $\nu<3$ mV/s (see details in SM Figure \ref{CV_PSD}).

\begin{figure}
\includegraphics[trim={11.5cm 0 12.5cm 0},clip,width=0.5\textwidth]{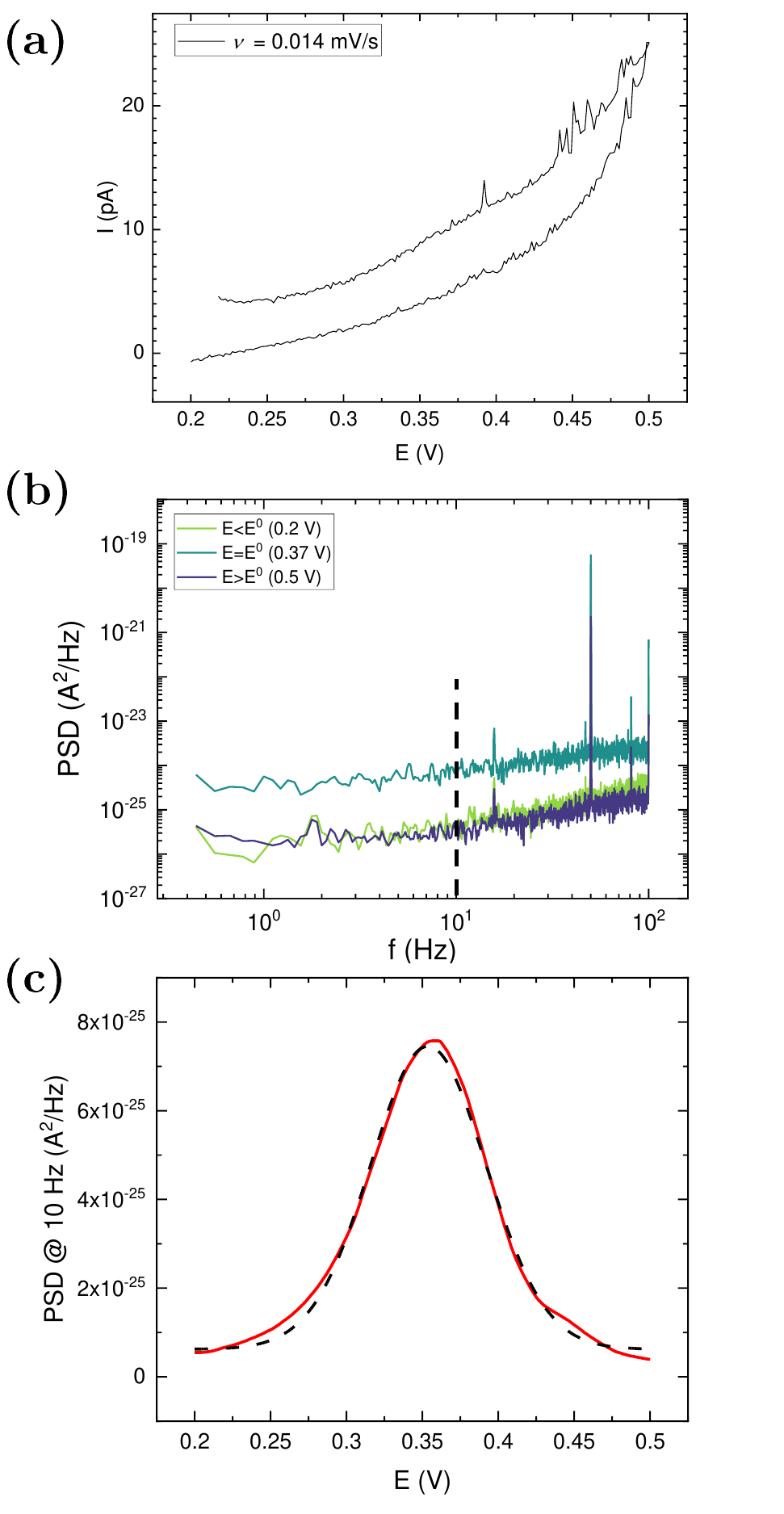}%
\caption{(a) CV of a FcC$_{11}$SH SAM on gold. (b) PSD measured at 0.014~mV/s at $E<E^0$, $E\approx E^0$ and $E>E^0$. (c) PSD versus potential at 10 Hz. The dashed line is fitted using Eq. \ref{S_f0_k} considering a lognormal distribution of $H$ with $\sigma=3.97\%$ of the mean value of $H$.\label{noise_spectrum}}
\end{figure}

Figure \ref{noise_spectrum} (a) shows a cyclic voltammetry (CV) scan at $\nu=0.014$ mV/s where no faradaic current signal can be identified. Figure \ref{noise_spectrum} (b) shows power spectral density (PSD) values at $E<E^0$, $E\approx E^0$ and $E>E^0$ on the forward scan. Figure \ref{noise_spectrum} (c) shows the variation of PSD at 10 Hz as a function of the voltage, showing a clear peak. These results demonstrate the ability to detect an electrochemical reaction at an electrode through shot noise measurements, even when the average current signal from the CV does not show any reaction. The number of molecules $N=7.5\times 10^{10}$ is calculated from the CV data at higher scan rates (Figure \ref{SM_sweeprates}), and using this value and Eq. \ref{S_V0_f0}, the peak amplitude of PSD data shown in Figure \ref{CV_PSD} yields $k_0 = 6.3\times10^{7}$~s$^{-1}$ ($z=1$~nm) is in good agreement with literature values for this molecule\cite{chidsey_free_1991,zevenbergen_fast_2009}.\\


The FWHM of the PSD peaks on Figure \ref{noise_spectrum} is $\approx 90$~mV, broader than the 56 mV predicted by Eq. \ref{FWHM_S}. Taking $\eta=0$ in the Eq. \ref{k_marcus} gives an expression for $k_0$, notably showing dependencies with $H$ and $\lambda$ \cite{smalley_kinetics_1995}. Previous work \cite{chidsey_free_1991} showed that variation of lambda within physically reasonable limits do not significantly impact the electron transfer rates. However, the electronic coupling term $H$ typically varies following lognormal distributions \cite{trasobares_estimation_2017} and can impact significantly the resulting value of $k_0$ with variations of just a few percent of its average value (see Figure \ref{k0_dist} \& \ref{PSD_dist}). If a lognormal distribution of the fluctuation of $H$ (and thus, of $k_0$ as well) is assumed, a standard deviation of $\sigma \approx 4\%$, comparable to what was reported in ref \cite{trasobares_estimation_2017} ($\approx 2\%$), can explain the broadening observed in PSD (Figure \ref{noise_spectrum} (c) dashed line).\\



There is a significant difference between molecular monolayers in liquid and solid-state devices in terms of electrostatic forces. In the first case, the electrostatic interactions between neighboring molecules are greatly reduced thanks to the high permittivity of water. Previous research on $\pi-\pi$ coulomb repulsion $\varphi$ (Eq. \ref{phi}) within similar Fc SAMs showed negligible impact on current CV \cite{trasobares_estimation_2017,reuter_signatures_2012}.

\begin{equation} \label{phi}
\varphi = q\frac{1-\left( 1+(\frac{r_a}{d})^2\right)^{-0.5}}{4 \pi \varepsilon_0 \varepsilon_r d}
\end{equation}

\noindent
with $d$ the intermolecular distance, $r_a$ the distance to counter ions, $\varepsilon_0$ the permittivity of the vacuum and $\varepsilon_r = \varepsilon_{H_2O}|\varepsilon_{SiO_2}$ the permittivity of the medium under consideration. Taking the same formalism and distribution of $d$ as in \cite{trasobares_estimation_2017} for $\varphi$ but changing $\varepsilon_{H_2O}=79$ to $\varepsilon_{SiO_2}=3.9$ results in variations of $E^S_{FWHM}$ from 1\% to 45\% respectively (see Figure \ref{phi_H2O} \& \ref{phi_SiO2}). As a result the screening of electrostatic interaction by water avoids a dispersion of the energy levels, such as the one observed in nanotransistors \cite{clement_one-by-one_2010}, and thus, avoids the domination of a $1/f$ noise resulting from the sum of multiple Lorentzian spectra with different amplitude/corner frequencies.

In conclusion, we demonstrated the measurement of the shot noise generated by an ensemble of surface-attached Fc redox molecules, which can be seen as identical single-electron boxes, in liquid and in ambient conditions. A formalism is proposed to understand it and exhibit dependencies between such noise and electronic coupling. This constitutes a further step toward nanoelectrochemistry and single molecule measurements, which could be practically achieved using our technique combined with a transducer such as a nanotransistor and be extended to other systems such as quantum dots monolayer \cite{fruhman_high-yield_2021}.

Our technique allows for the measurement of electron transfer rates at low frequencies without the need for highly time-resolved instrumentation. Although we compared our technique with traditional voltammetry techniques, exhibiting a clear signal in PSD when $I$ tended to zero, the very concept of ``potential scan'' is actually not required to perform noise measurements. As few as two points at potentials far from $E^0$ and one at $E^0$ can suffice to resolve the eventual background noise of the experiment and the noise due to the attached molecule, yielding $k_0$ and $N$ provided the knowledge of $\beta$ and $z$. Concurrently, since the measurements is carried out at equilibrium, capacitive contributions are altogether avoided, improving the signal and simplifying drastically the interpretation of the data. This opens perspectives in the field of biosensors \cite{li_redox-labelled_2022}, where the limit of detection of existing techniques could be further extended by shot noise analysis, and in high-frequency molecular diodes, where the electron transfer rate can be estimated through the low-frequency noise.

\section{Acknowledgments}
\begin{acknowledgments}
This work has been supported by the EU-ATTRACT project (Unicorn-Dx), the French ``Agence Nationale  de la Recherche''(ANR) through the ``SIBI'' project (ANR-19-CE42-0011-01) and the JSPS Core-to-Core Program (JPJSCCA20190006).\\

S.G designed the acquisition system, conducted the experiments and data analysis and developed the theory, S.L. fabricated the devices, L.J. designed the acquisition system, SH. K. and A.C. contributed to the scientific interactions on electrochemistry, C.D. and N.C. conceived and supervised the whole project. All authors actively contributed to the discussions and the writing of the paper.
\end{acknowledgments}


\bibliography{Paper_noise_electrochemistry}

\end{document}



\title{Electrochemical Shot Noise of a Redox Monolayer\\
			Suppplementary Material}


\author{Simon Grall, Shuo Li, Laurent Jalabert, Soo-Hyeon Kim}
\affiliation{IIS, LIMMS/CNRS-IIS IRL2820, The Univ. of Tokyo; 4-6-1 Komaba, Meguro-ku Tokyo, 153-8505, Japan}
\author{Arnaud Chovin, Christophe Demaille}
\affiliation{Université Paris Cité, CNRS, Laboratoire d'Electrochimie Moléculaire, F-75013 Paris, France}
\author{Nicolas Cl\'ement}
\affiliation{IIS, LIMMS/CNRS-IIS IRL2820, The Univ. of Tokyo; 4-6-1 Komaba, Meguro-ku Tokyo, 153-8505, Japan}


\date{\today}


\maketitle
\beginsupplement
%

%

\begin{figure}
\includegraphics[trim={8cm 2cm 8cm 0},clip,width=\textwidth]{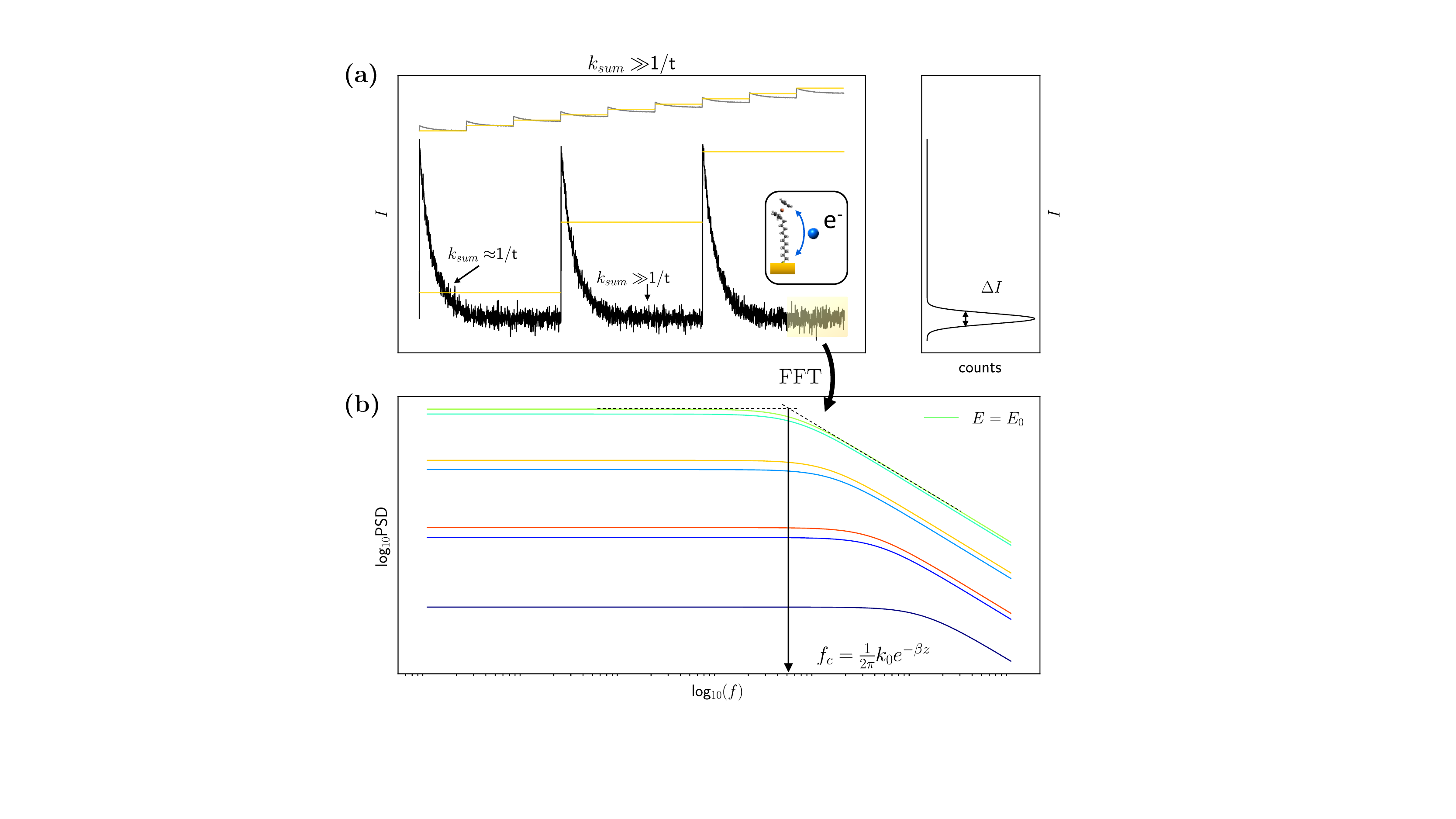}%
\caption{(a) Computer-generated time traces of the current for illustrative purposes, with long waiting time such as $I$ reaches zero at each voltage step (black). Voltage steps $E_{step}$ are represented in yellow. The right panel represents the distribution of the current around its average, with $\Delta I$ as described in the main text.  (b) Noise versus frequency for different $E$, with the corner frequency indicated for $E=E^0$ The noise data are obtained at each $E_{step}$ by Fourier transform of the autocorrelation of the current data when the average current reaches zero (yellow window on (a)). 
\label{SM_FFT}}
\end{figure}

\clearpage
\section{Sample preparation}

\begin{figure}
\includegraphics[trim={0cm 0 3cm 0},clip,width=\textwidth]{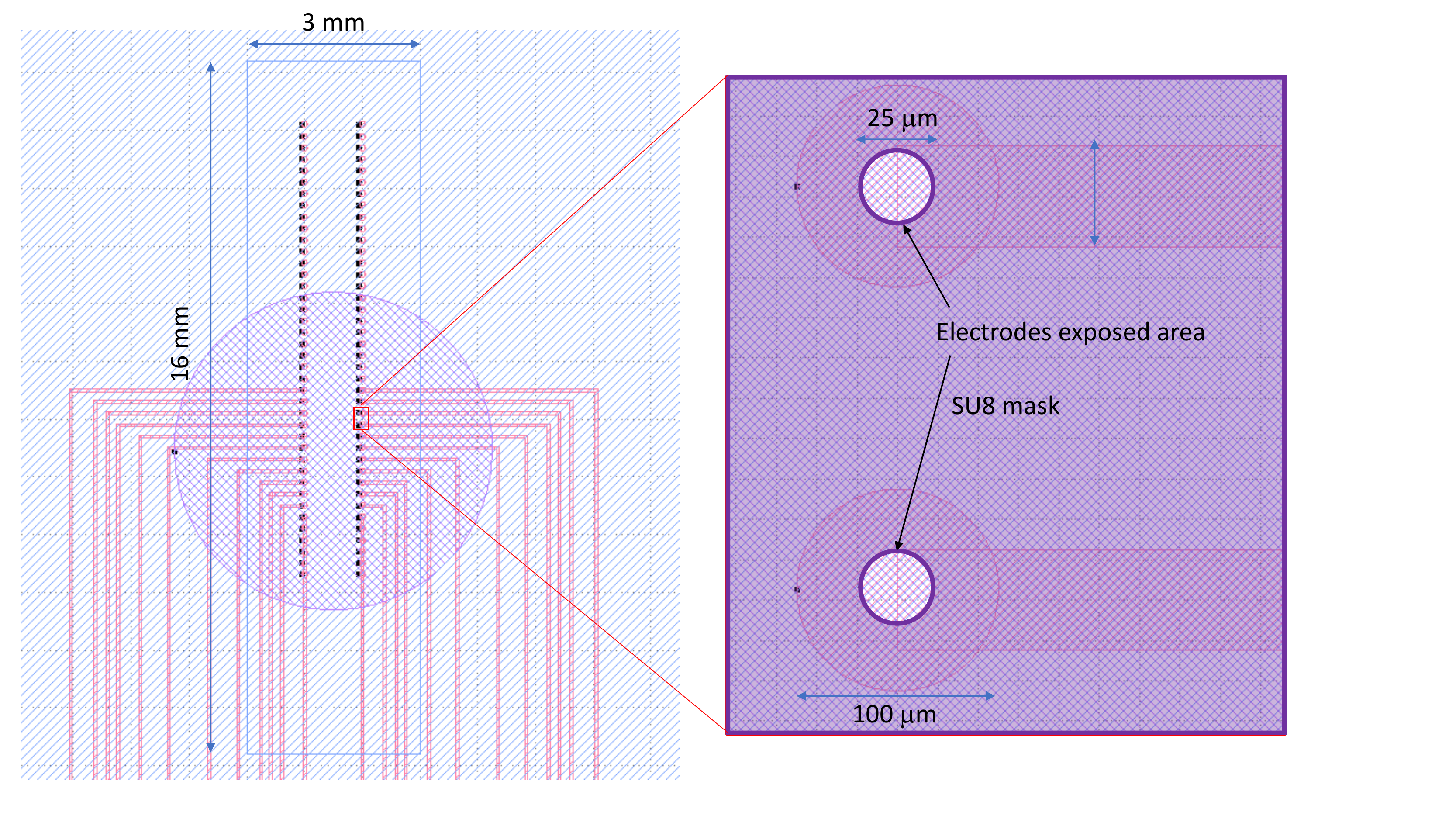}%
\caption{Layout of the microelectrodes used, with a zoom on the right indicating the exposed area and the SU8 mask.\label{map_electrodes}}
\end{figure}

The electrode is fabricated by gold sputtering on a silicon wafer. Plain electrodes (From 45 mm$^2$ down to 0.78 mm$^2$) and microelectrodes (as described Figure \ref{map_electrodes}) were used. The later was preferred for noise measurements, as in a 2-electrode electrochemical setup, keeping low-current to avoid potential drop is critical. For the microelectrodes configuration, a layer of SU8 (thickness = 25 \textmu m) is spin-coated on all the wafer except the designated gold areas to prevent unwanted reactions with silicon dioxide and limit the parasitic capacitance (Figure \ref{map_electrodes}). The electrode is incubated in a 1 mM ethanol solution of ferrocene undecanethiol  for at least one day, with further incubation in a 1 mM undecanethiol solution for 2h. CV curves Figure \ref{SM_sweeprates} were used to extract $N= 7.5\pm 0.05\times10^{10}$.
The electrochemical experiments are all carried out in 0.1 M NaClO$_4$ aqueous electrolyte and after N$_2$ bubbling. The electrochemical cell (SEC-3F, ALS Co., Japan) is made of a small silicon gasket delimiting a channel ($3\times 16 \times 0.3$ mm) over the microelectrodes and connected to the Ag/AgCl electrode. The cell can be sealed with microfluidic switches after filling with the electrolyte.

\clearpage
\section{Measurement setup}
\begin{figure}
\includegraphics[trim={1cm 0 1cm 0},clip,width=\textwidth]{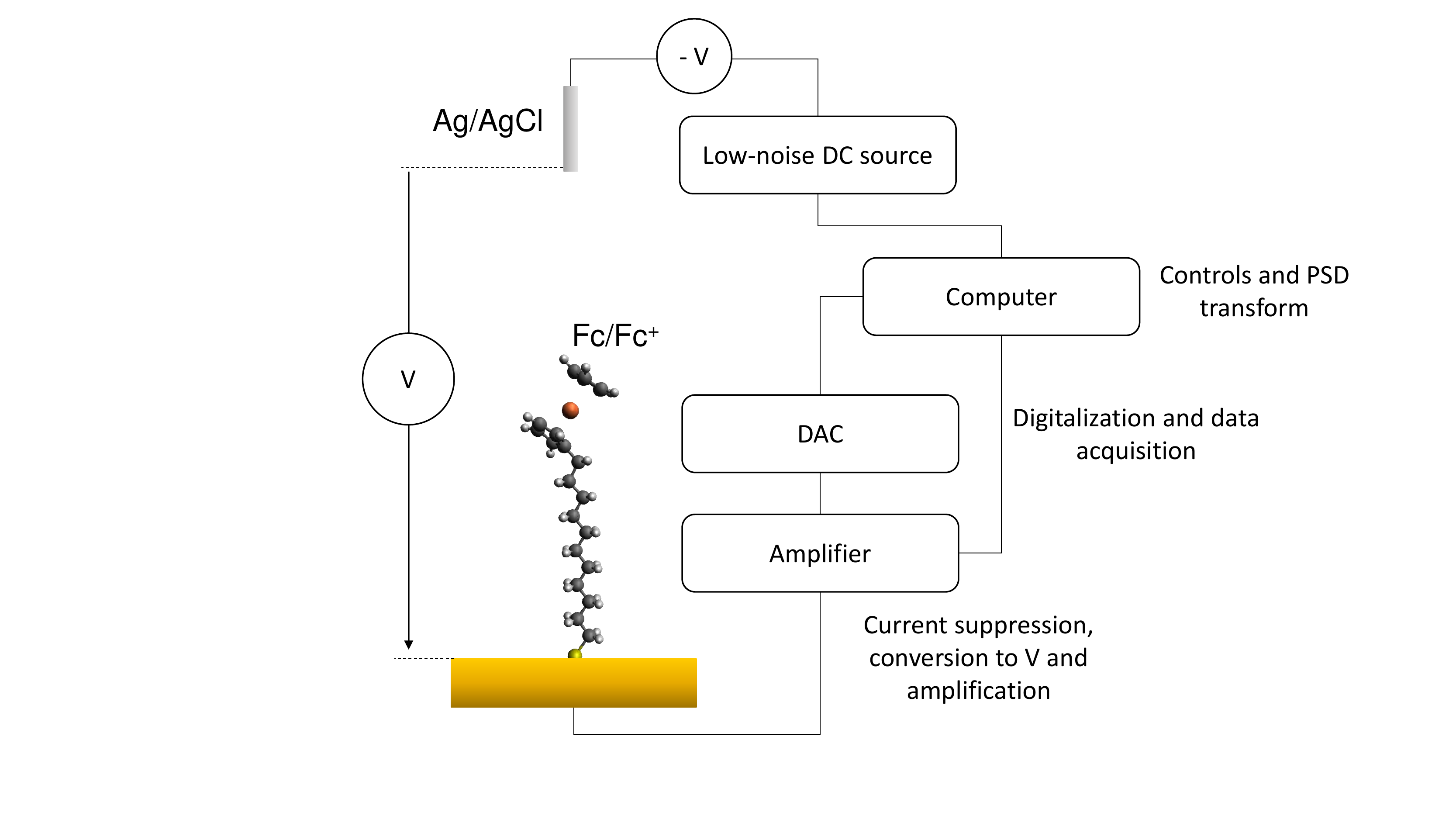}%
\caption{Schematics of the setup used to scan the voltage over the Fc SAM and recover the current and PSD data. \label{SM_measurement_setup}}
\end{figure}

The microelectrodes are connected to a current amplifier (CA5351, NF Corporation, Japan), itself connected to a digital-to-analog converter (USB-4431, National Instrument) (Figure \ref{SM_measurement_setup}). The voltage is applied using a Ag/AgCl (3 M NaCl) reference electrode. Though only two electrodes are used, the very low currents expected make the voltage drop across the electrochemical cell negligible. As a control, we verified that experiments carried out in a three-electrode configuration (using a platinum wire counter electrode) yielded similar results as those obtained with our two-electrode setup. The whole experiment is controlled using a Labview program. As a precaution, a platinum counter electrode is added to the setup and cyclic voltammograms are measured both with and without the counter electrode, with no significant differences.

\clearpage

\begin{figure}
\includegraphics[trim={7cm 1cm 6cm 0},clip,width=\textwidth]{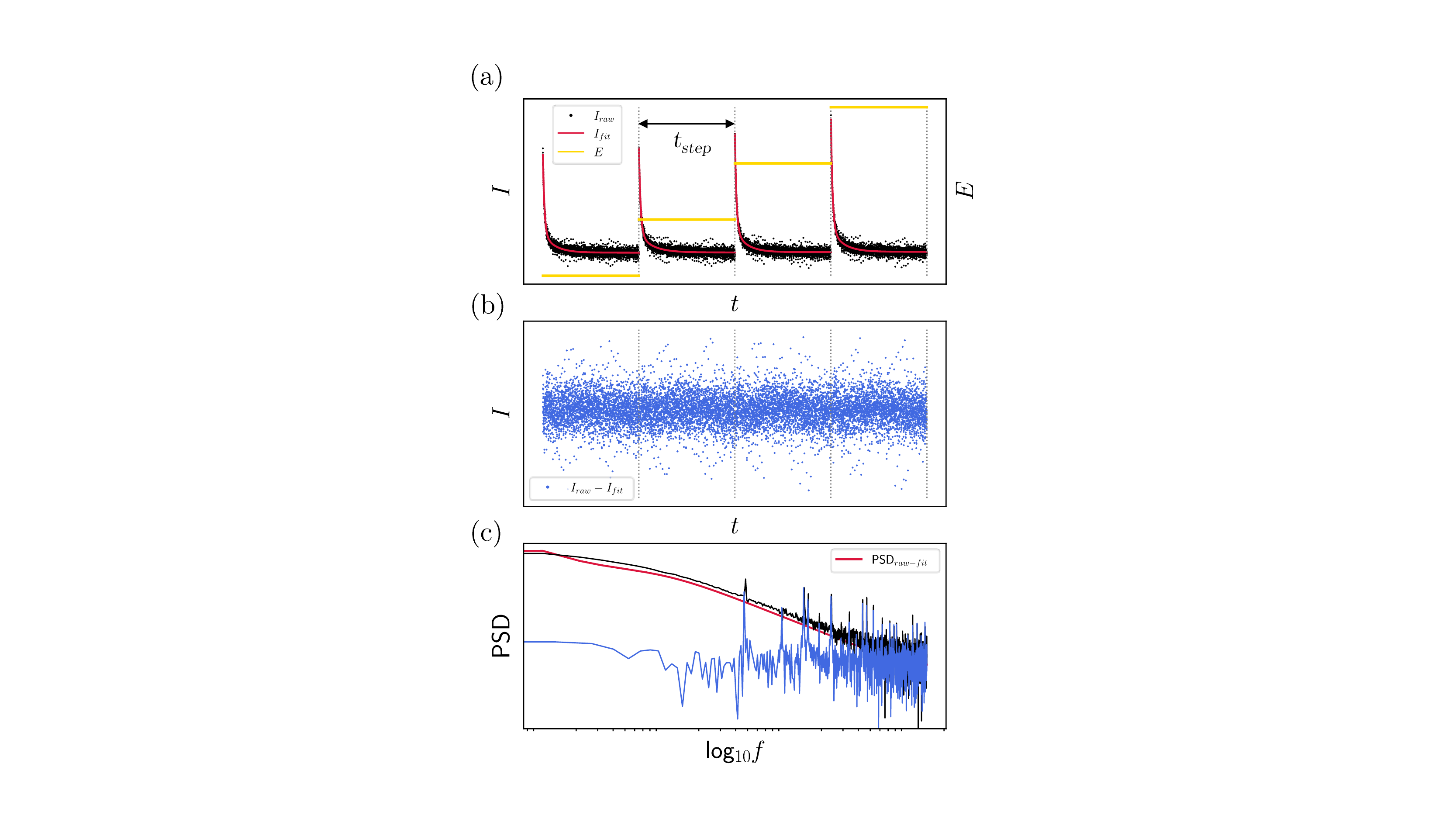}%
\caption{(a) Staircase voltammetry example, with the raw current data (black dots), a double exponential decay fit of the current (red) and the voltage steps (yellow). (b) Flattened currents (raw current corrected by the subtraction of exponential fits). (c) PSD spectrum of one current time trace obtained using the raw current subtracted with the exponential fit (blue). The black line represents the PSD obtained from raw data, and the red one from the exponential fits. \label{SM_timetrace_trimming}}
\end{figure}

\begin{figure}
\includegraphics[trim={7cm 0 6cm 0},clip,width=\textwidth]{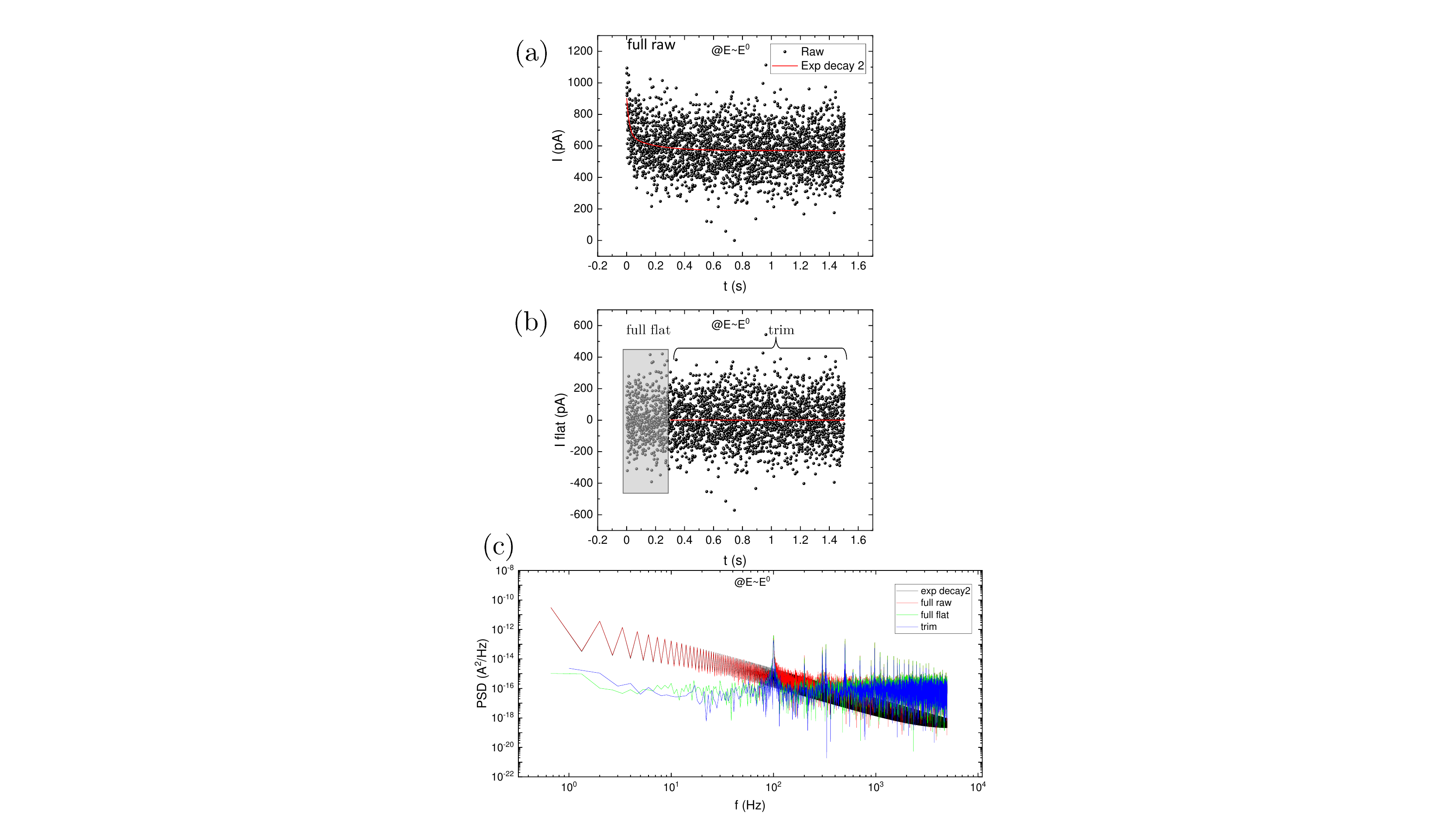}%
\caption{(a) Raw current data trace, a double exponential decay fit of the current is shown (red). (b) Flattened current (raw currents corrected by the subtraction of exponential fits) with only the last 4/5$^{th}$ kept for PSD calculation. (c) PSD spectra obtained from raw data (black), exponential fits (red), flattened current (green) and trimmed flattened current (blue). The later is used throughout this work for processing "CV conditions" data. \label{SM_trimming}}
\end{figure}

\section{Experimental measurements conditions}

\begin{figure}
\includegraphics[trim={4cm 5cm 5cm 4cm},clip,width=1\textwidth]{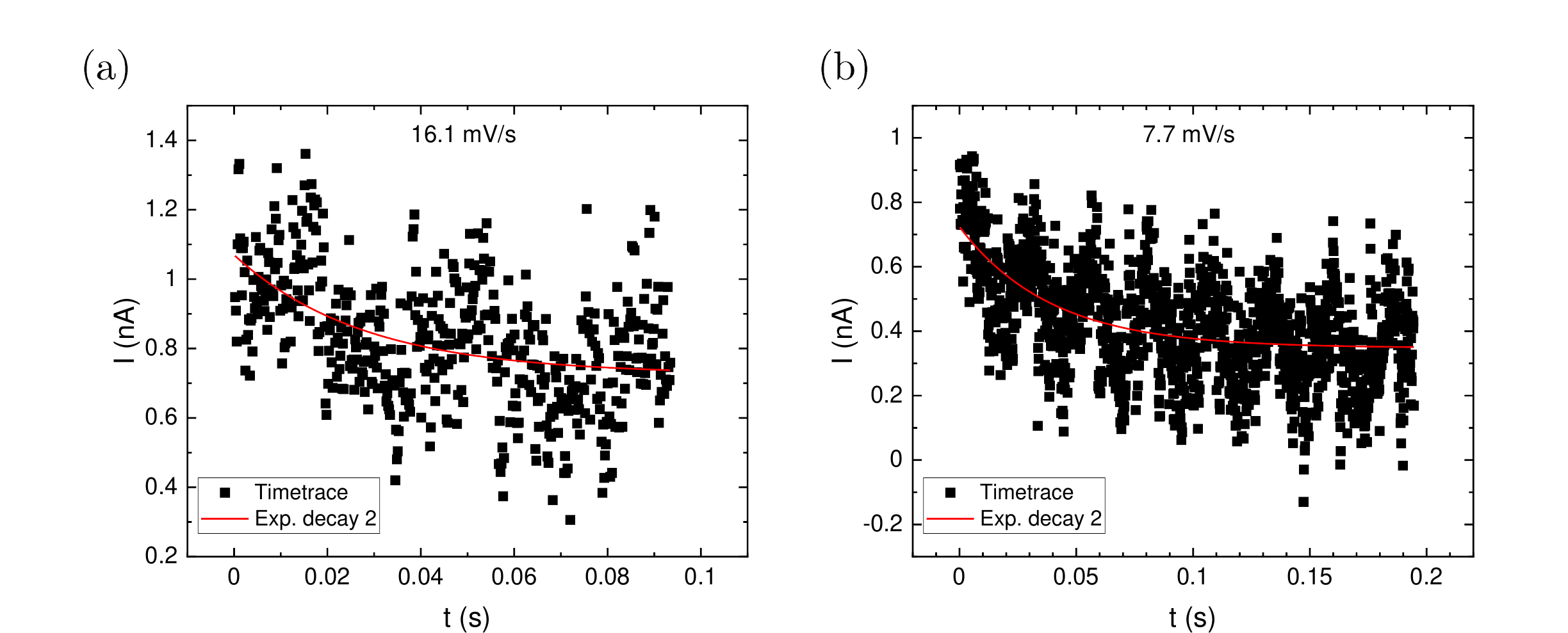}%
\caption{time traces recorded at $E=E^0$ at (a) 16.1 mV/s and (b) 7.7 mV/s. The fitting lines are obtained using Eq. \ref{exp_fit} \label{SM_timetraces}}
\end{figure}

To avoid the contribution of any transient current, two types of measurements are realized. The first one corresponds to staircase sampled current experimental conditions (``SCV conditions''). For each potential step $E_{step}$ the whole timetrace of the current is recorded and fitted with a double exponential decay (Figure \ref{figure1} (c), details in Figure \ref{SM_trimming} and \ref{SM_timetraces}):

\begin{equation}\label{exp_fit}
I=A_ce^{\frac{-t}{\tau_c}}+A_fe^{\frac{-t}{\tau_f}}
\end{equation}

\noindent 
with $(A_c,\tau_c)$ the amplitude and time constant of the capacitive current, due to the relaxation of the double layer, and $(A_f,\tau_f)$ the amplitude and time constant of the faradaic current, due to the charge transfer between Fc molecules and the electrode. Typical time constants were on the order of 0.1 s and 0.01 s for $\tau_f$ and $\tau_c$ respectively.  The PSD spectra are extracted from ``flattened'' currents, obtained by subtracting the exponential decay contribution (Figure \ref{figure1} (d), details in Figure \ref{SM_FFT} $-$ \ref{SM_timetraces}). For $t_{step} > 5\times max(\tau_c,\tau_f)$ (here corresponding to $\nu <10$~mV/s), the transients were well-resolved and the low-frequency noise could be acquired using this method (Figure \ref{figure1} (e)). For $\nu< 10$~mV/s, the time trace was fitted, straightened and trimmed, keeping only the $4/5^{th}$ of the time trace to calculate the PSD (Figure \ref{SM_timetrace_trimming}). The resulting PSD does not vary significantly as shown in Figure \ref{CV_PSD}, which supports $t>5\times \tau$ criteria being reasonable.\\

\begin{figure}
\includegraphics[trim={8cm 0cm 8cm 0cm},clip,width=\textwidth]{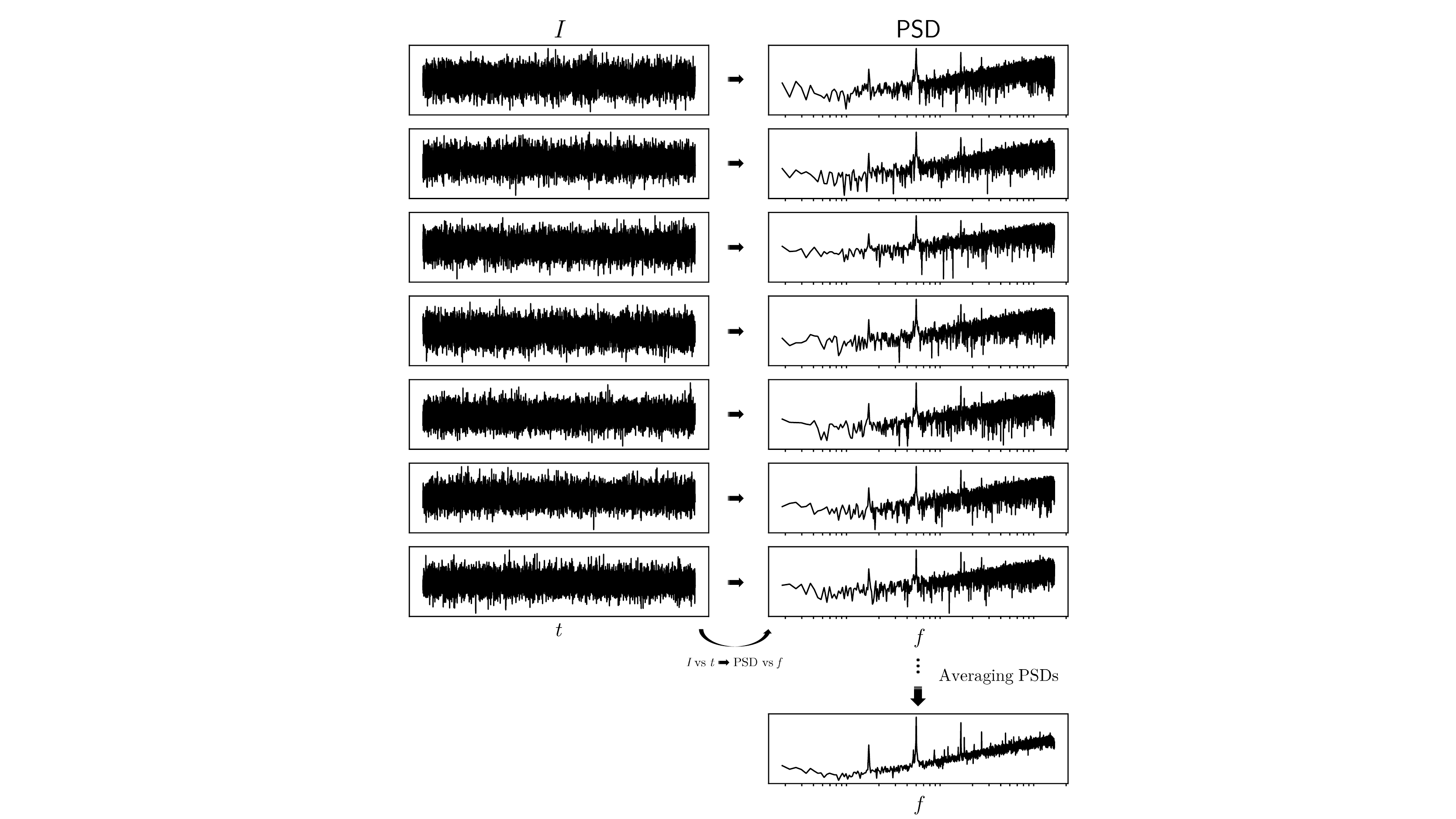}%
\caption{Illustration for the "Noise conditions" measurements. For each voltage step, the current is measured $n$ times ($n=7$ in this example, $n>10$ during experiments) during a time window of $t$ and converted into PSD (plotted versus the frequency $f$). After $n$ measurements, the PSDs obtained are averaged over $n$ (bottom right graph) and the measurement restarts at the next voltage step. \label{SM_averaging}}
\end{figure}

\subsection{Noise conditions}
The second type of measurement, called here ``Noise conditions'' consists in using very slow scan rates ($\nu < 0.1$~mV/s) in order to bring the system as close as possible to equilibrium at each electrode potential value (Figure \ref{SM_FFT}). Several time traces are recorded at each voltage step, giving time to the system to reach equilibrium for a given voltage ($k_{sum}\gg 1/t$), meaning that after the initial time trace recorded just after the voltage change, no transient phenomenon occurs. In this case, the mean current tends to zero (Figure \ref{figure1} (a) and (c)), but not the noise (Figure \ref{figure1} (b) and \ref{SM_FFT}). This allows to average the PSD obtained from each time trace at least 10 times (Figure \ref{SM_averaging}), and this for each voltage step with no transient contribution to the PSD. The other consequence is that the effective voltage scan rates are very low, on the order of 0.1 mV/s and below. The measure of the noise out of equilibrium (if it can still be called "noise" in such case) is possible but is beyond the scope of the present article.\\

In Figure \ref{CV_PSD}, PSD measurements were carried out from $\nu=$ 1 mV/s up to  16.1 mV/s only, due to transients interfering at higher scan rates. Lower scan rates were measured using the Noise conditions.

%
%

\clearpage

\section{Cyclic voltammetry currents}
The probability $P_{ox}$(E) for a molecule to be oxidized at a given voltage $E$ after time $t$ can be written as :

\begin{equation} \label{pox}
P_{ox}=\frac{k^{BV}_{ox}}{k_{sum}}(1-e^{-(k_{sum}t_{step})})
\end{equation}

with $k_{sum}=k^{BV}_{ox}+k^{BV}_{red}$, $\nu=E_{step}/t_{step}$ the voltage scan rate. In the case where the electron transfer is fast compared to the time spent at each step ($k_{sum}\gg 1/t_{step}$), $P_{ox}$ simplifies to:

\begin{equation} \label{pox_eq}
P_{ox}=\frac{k^{BV}_{ox}}{k_{sum}}=\frac{1}{1+e^{-\frac{\eta}{k_BT}}}
\end{equation}

The proportion of attached molecules being effectively oxidized at a given time is approaching $P_{ox}$ following an exponential decay as all molecules end up in the equilibrium state. Defining the current $I$ as the number of transferred charges per unit of time, the current is:

\begin{align} 
I =& Nq \frac{dP_{ox}}{dt_{step}} =Nq\nu \frac{dP_{ox}}{dE_{step}} \label{I}
\end{align}

\noindent
with $N$ the total number of molecules. For the rest of the study, unless mentioned otherwise, we consider the case of relatively slow scan rates, where electron transfer rates are large compared to the inverse of the time spent at each voltage step ($k_{sum}\gg 1/t_{step}$). $t$ is defined here as the sampling time, with $0\leq t\leq t_{step}$. In this case, for long sampling times (i.e. $k_{sum}\gg 1/t$), $P_{ox}=k_{ox}/k_{sum}$ and $I$ simplifies to:

\begin{equation} \label{I_simple}
I =  \frac{N q\nu}{4k_BT}\frac{1}{\cosh^2(\frac{\eta}{2k_BT})}	
\end{equation}
with a full width at half maximum (FWHM):

\begin{equation} \label{FWHM_I}
E^I_{FWHM} =  4\operatorname{acosh}(\sqrt{2})\frac{k_BT}{q}\approx 90.6 \text{~mV}
\end{equation}
where $\operatorname{acosh}$ is the inverse of the hyperbolic cosine (considering $T=$ 298 K for the numerical value) \cite{laviron_general_1979,huang_random_2013}. \\

\section{Dielectric losses}
\begin{figure}
\includegraphics[trim={8cm 5cm 9cm 3cm},clip,width=0.8\textwidth]{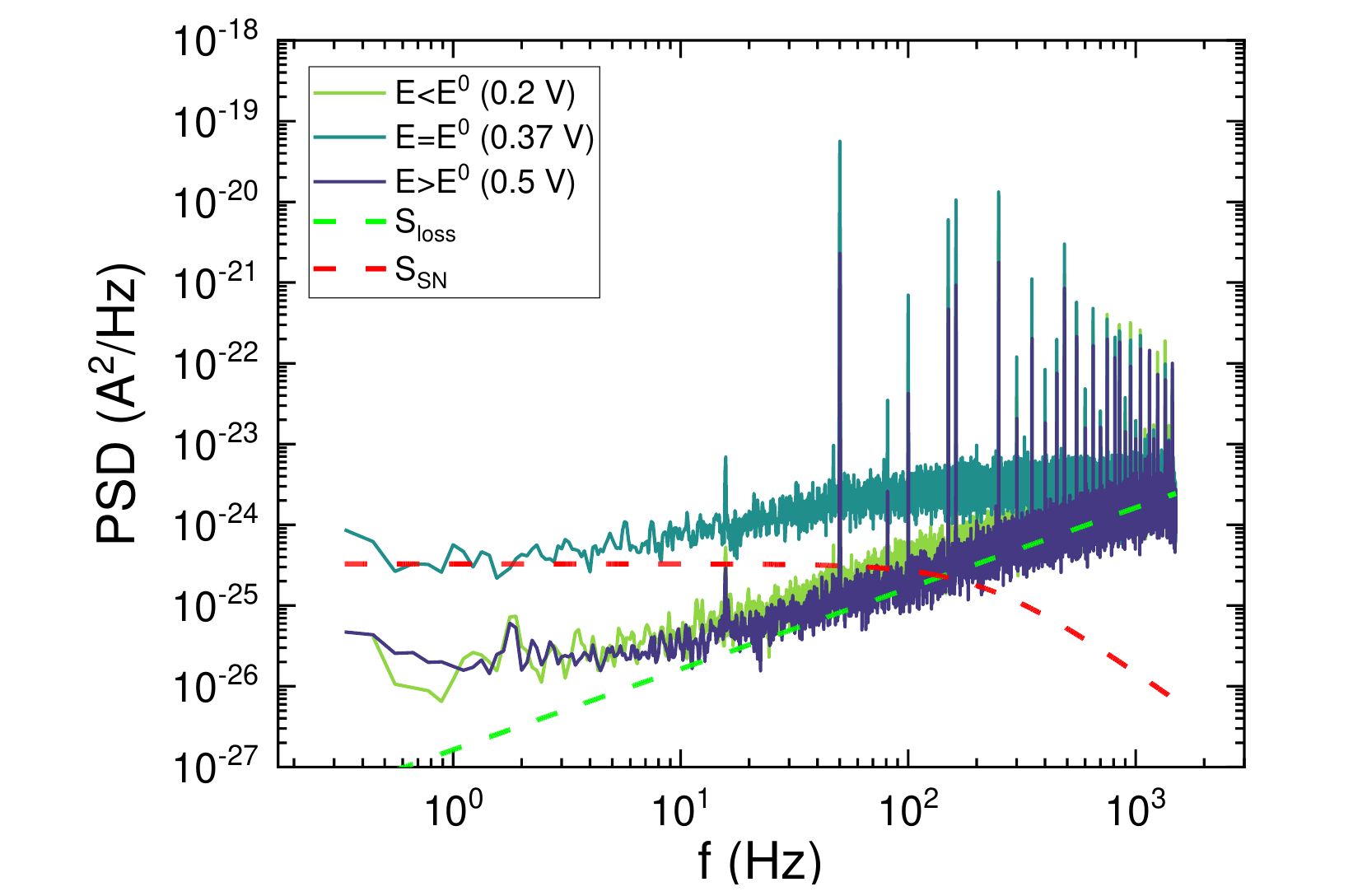}%
\caption{Full PSD spectrum with partial fits (dashed lines) of dielectric noise  losses (red) at $E\ll E^0$ and electrochemical shot noise (green) at $E=E^0$. \label{SM_PSD_full_spectrum}}
\end{figure}

The PSD spectra exhibit a $\tan(\delta)$ dielectric loss contribution:

\begin{equation} \label{S_dielectric}
S_D(f) = 8k_BT\pi C\tan(\delta)f
\end{equation}
\noindent
with $C$ the capacitance and $\tan(\delta)$ the dielectric loss tangent \cite{westerlund_capacitor_1994,wen_generalized_2017}. Negligible at low-frequency, the dielectric noise allows access to the low-frequency shot noise, but unfortunately not to $f_c$.
The origin of the dielectric losses noise can be attributed to the SAM's capacitance. Considering the area of the exposed electrode ($A=7.84\times10^{-9}$m$^2$) and the capacitance of the SAM ($C=1.5\times10^{-8}$ F), PSD acquired far from $V_0$ can be fit with Eq. \ref{S_dielectric} (Figure \ref{SM_PSD_full_spectrum}) with dielectric losses around 1, slightly higher than measured in dry \cite{clement_relaxation_2010} and on the order of common polymer \cite{prabha_study_2015}. However, this assumes a constant $\tan(\delta)$ with frequency, which is not necessarily the case if for example, water molecules penetrate the SAM. The curve obtained at $E^0$ Figure \ref{SM_PSD_full_spectrum} seems to suggest a non-constant loss with frequency. Though beyond the scope of the current work, further investigations could allow to use such noise measurements as a probing method for the effective capacitance of the SAM\cite{israeloff_dielectric_1996}. The very small scaling of the PSD peak value  with $\nu$ confirms that the measurement of the noise is well-suited for the study of systems with small numbers of charge carriers, high electron transfer rates and at very slow scan rates, where the parasitic capacitive currents can be suppressed.

\clearpage

\begin{figure}[h]
\includegraphics[trim={11cm 0cm 11cm 0cm},clip,width=0.7\textwidth]{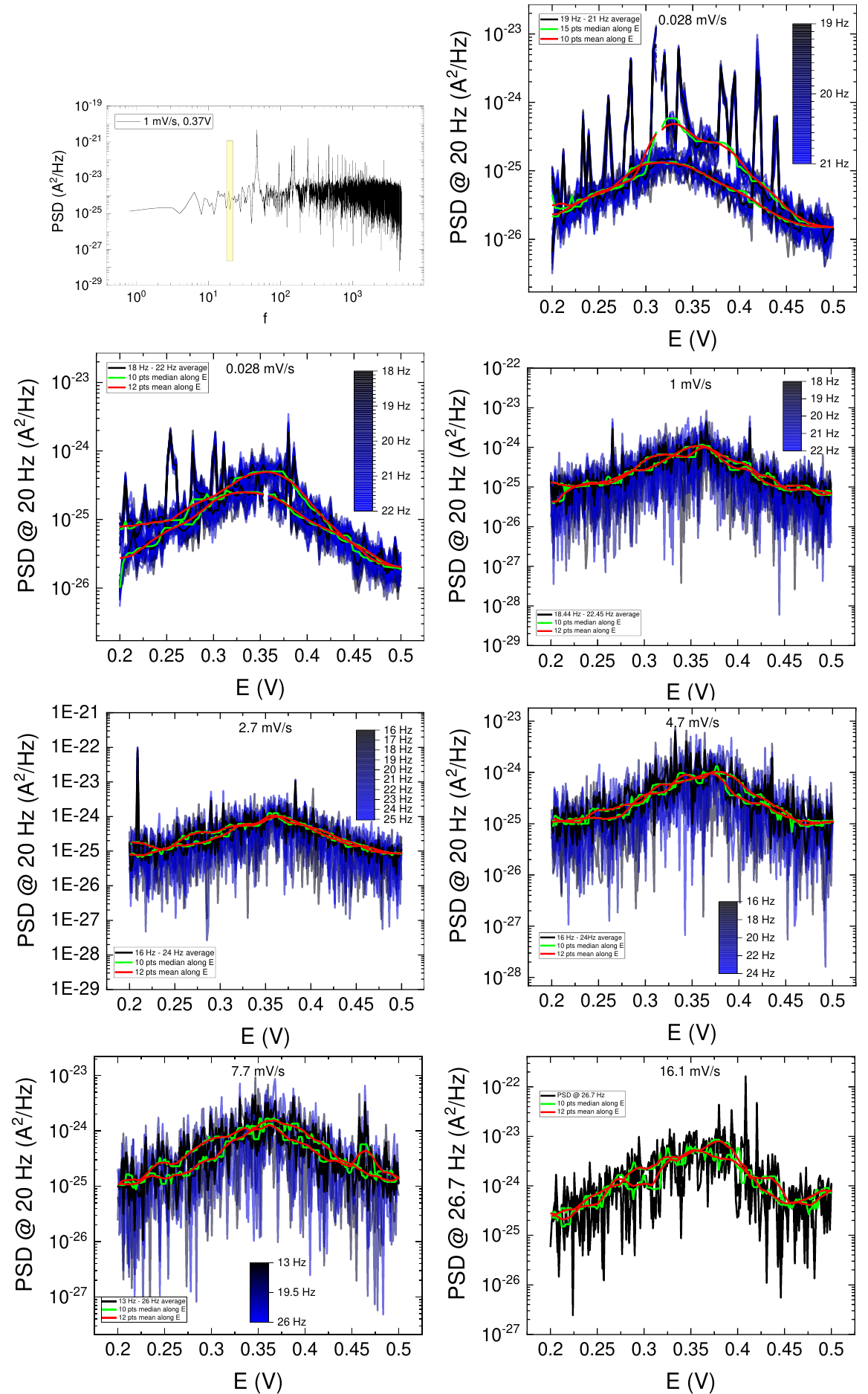}%
\caption{Smoothing procedure used to obtain Figure \ref{CV_PSD} in the main text. The noise is taken at 20 Hz (yellow square on the spectrum), averaging with neighboring frequencies (blue lines, average in black).   A rolling median (window$\approx$ 7\%, green line) followed by an adjacent averaging (window$\approx$7\%, red line) were used.\label{SM_sweeprates_smoothing}}
\end{figure}

\begin{figure}
\includegraphics[trim={11cm 0 12cm 0},clip,width=0.6\textwidth]{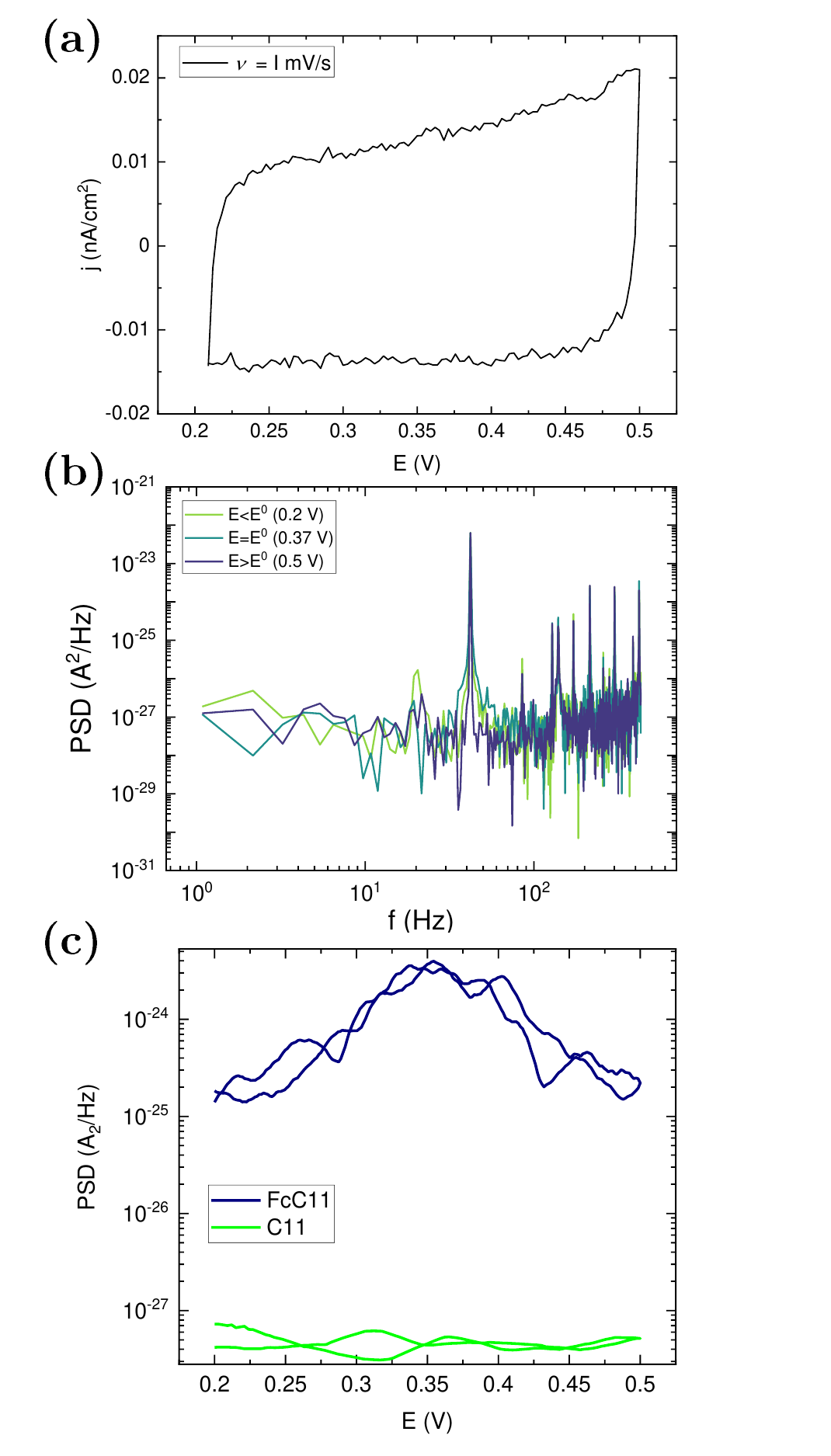}%
\caption{(a) CV of a C$_{11}$SH SAM on gold. (b) PSD measured at 1~mV/s at $E<E^0$, $E\approx E^0$ and $E>E^0$. (c) PSD versus potential at 10 Hz, comparing C$_{11}$SH and FcC$_{11}$SH curves obtained at the same scanrate (1~mV/s).\label{SM_C11}}
\end{figure}

\begin{figure}[h]
\includegraphics[trim={4cm 0 4cm 0},clip,width=\textwidth]{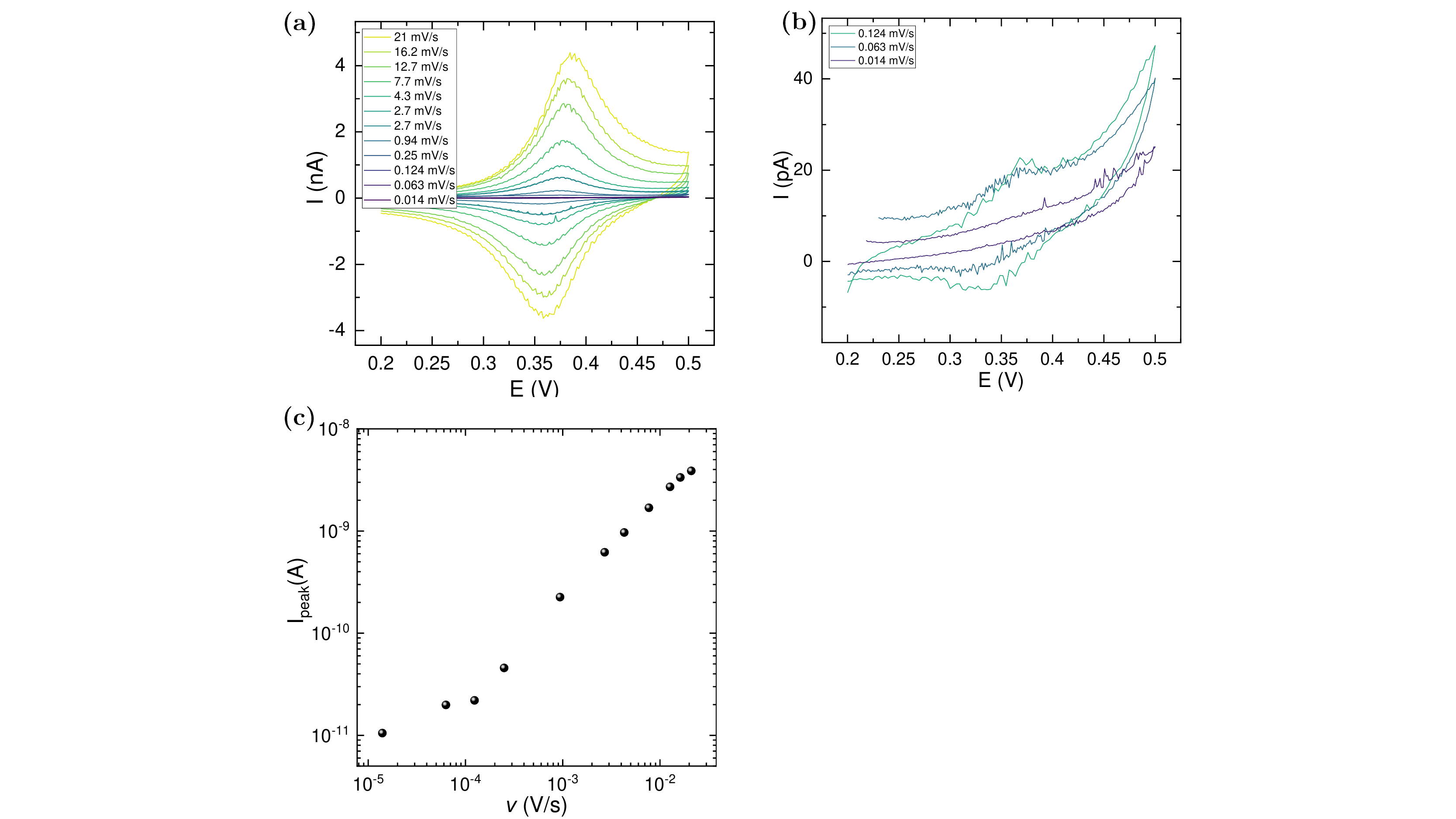}%
\caption{Current CV (a) with zoomed data (b) corresponding to the data shown Figure \ref{CV_PSD} in the main paper. (c) shows the dependency of the current with the scan rate. From the area of the oxidation peak on the current CV, we estimate the number of molecules $N \approx 7.5\times10^{10}$. \label{SM_sweeprates}}
\end{figure}

\begin{figure}[h]
\includegraphics[trim={4cm 5cm 4cm 2cm},clip,width=\textwidth]{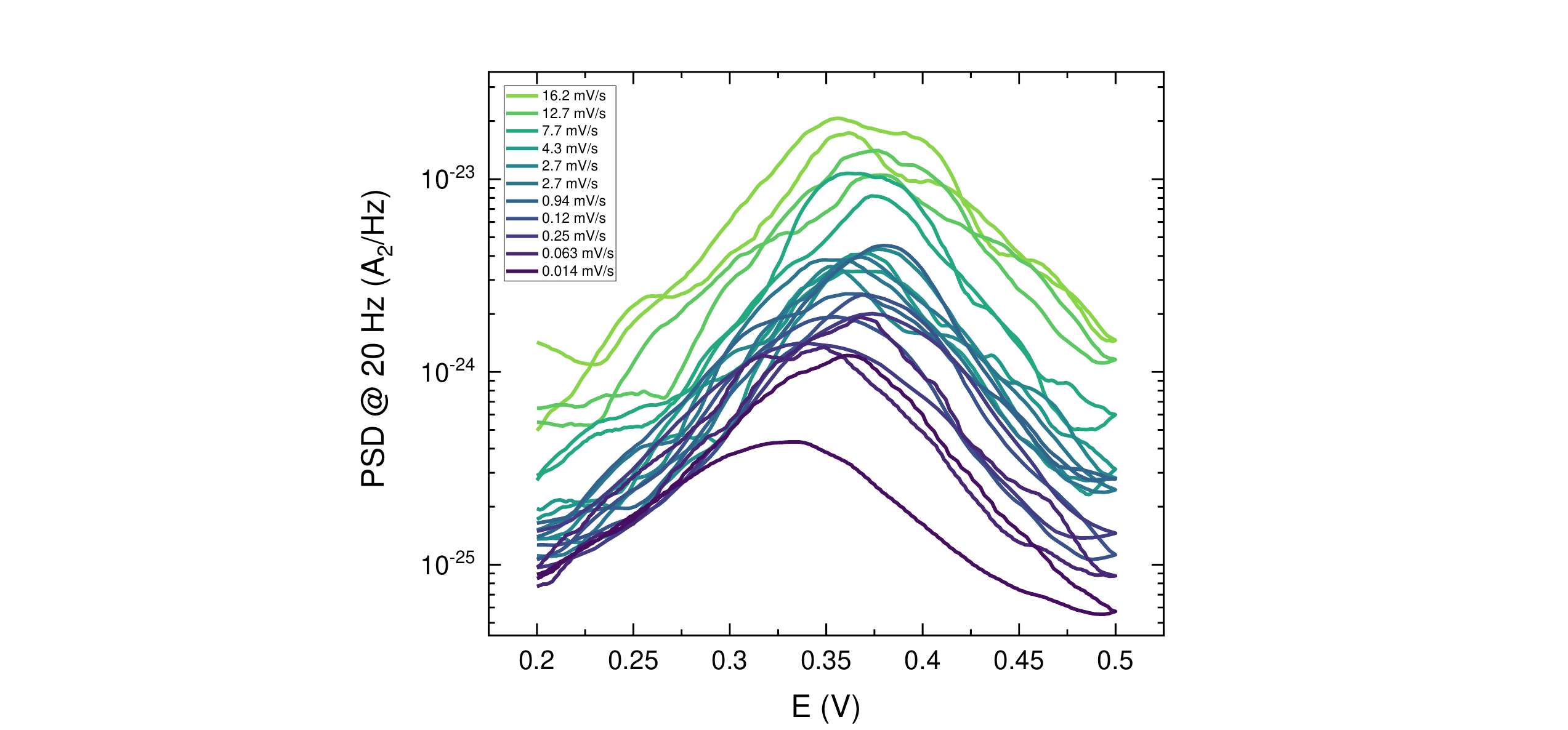}%
\caption{PSD of the current versus $E$ obtained at different $\nu$, at $f \approx$ 20 Hz, corresponding to the data shown in the main paper Figure \ref{CV_PSD}. \label{SM_sweeprates_PSD}}
\end{figure}
\clearpage

\section{Fluctuations of $k_0$}
\begin{figure}[h]
\includegraphics[trim={0cm 0 0cm 0},clip,width=\textwidth]{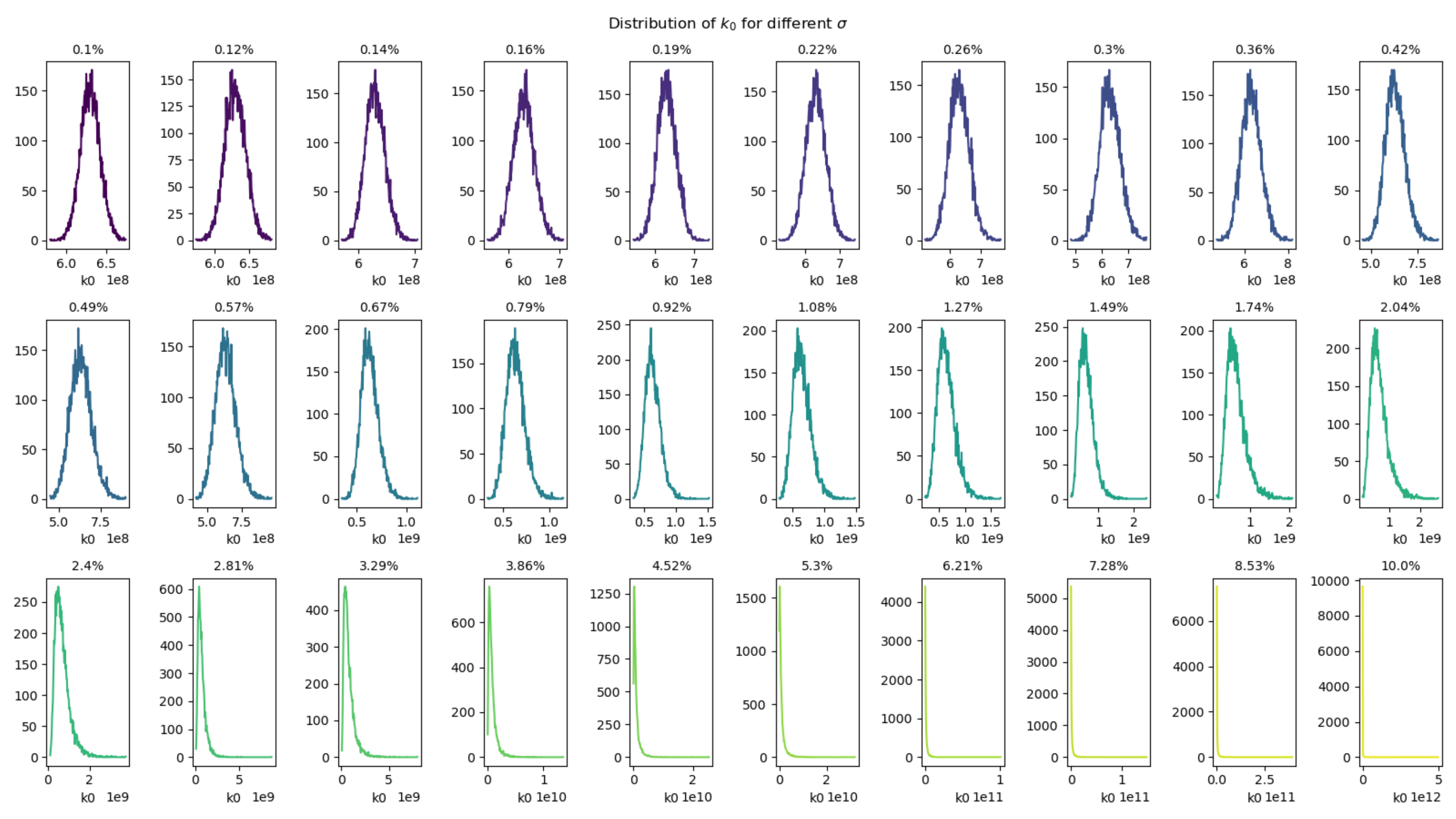}%
\caption{Distribution of $k_0$ resulting from fluctuations of the electronic coupling (percentages indicated above each graph). \label{k0_dist}}
\end{figure}

\begin{figure}[h]
\includegraphics[trim={0cm 0 0cm 0},clip,width=\textwidth]{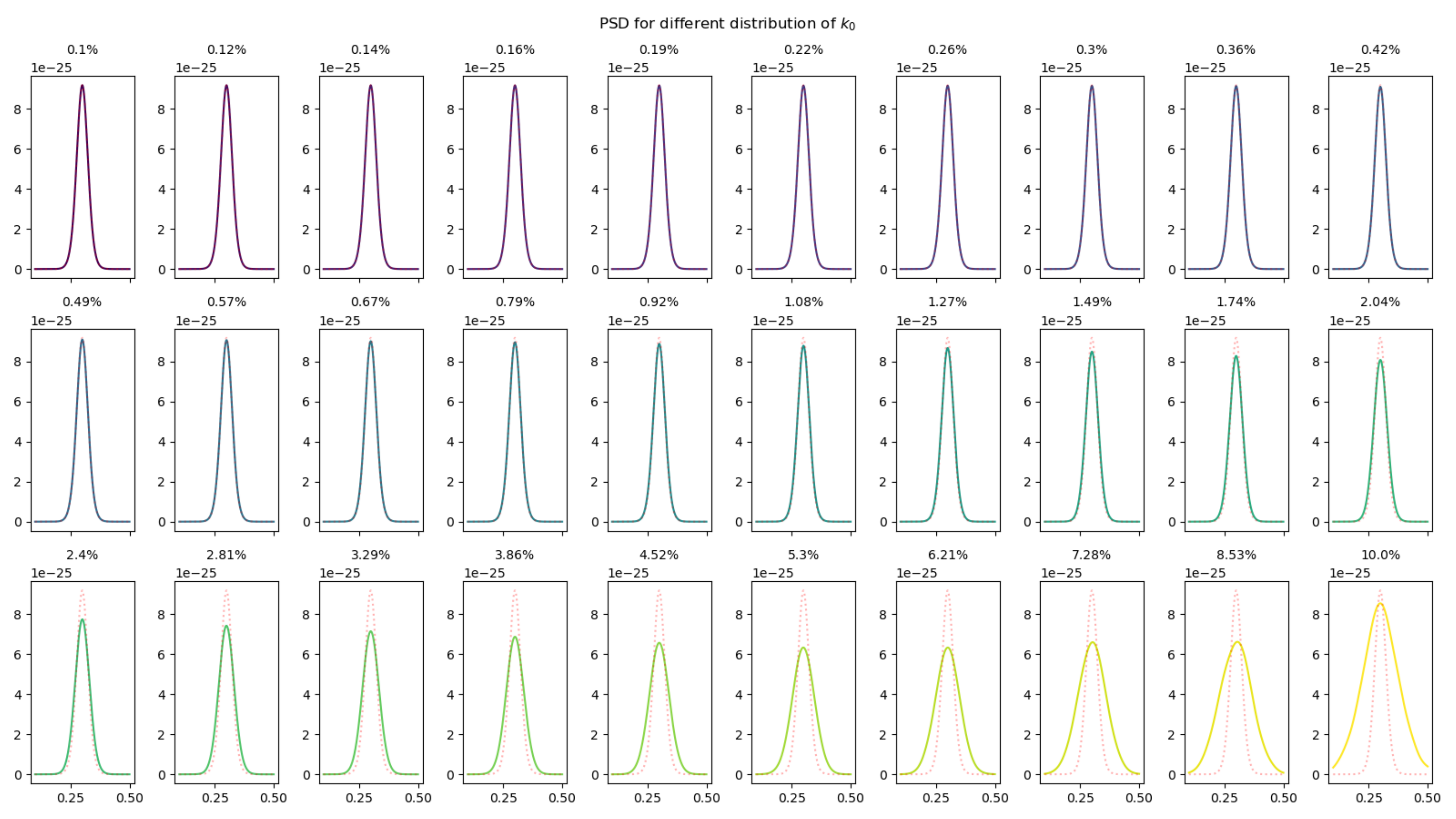}%
\caption{PSD voltnoisograms resulting from fluctuations of the electronic coupling (percentages indicated above each graph). The dashed line is the PSD obtained with no fluctuations of $k_0$.\label{PSD_dist}}
\end{figure}

Based on the simple Butler-Volmer description of the electron transfer, we assume a lognormal distribution of the electronic coupling $H$ such as $k_0\propto H^2$. $H$ is taken so that the average value of $6.3\times$10$^8$ s$^{-1}$ for $k_0$, with a $\sigma$ from 0.1\% to 10\%, similar to what was observed in \cite{trasobares_estimation_2017}. This leads to the following distributions of $k_0$ Figure \ref{k0_dist}, with associated noise for PSD Figure \ref{PSD_dist}.

\clearpage

\section{Impact of the permittivity of water}
\begin{figure}[h]
\includegraphics[trim={4cm 0 5cm 0},clip,width=\textwidth]{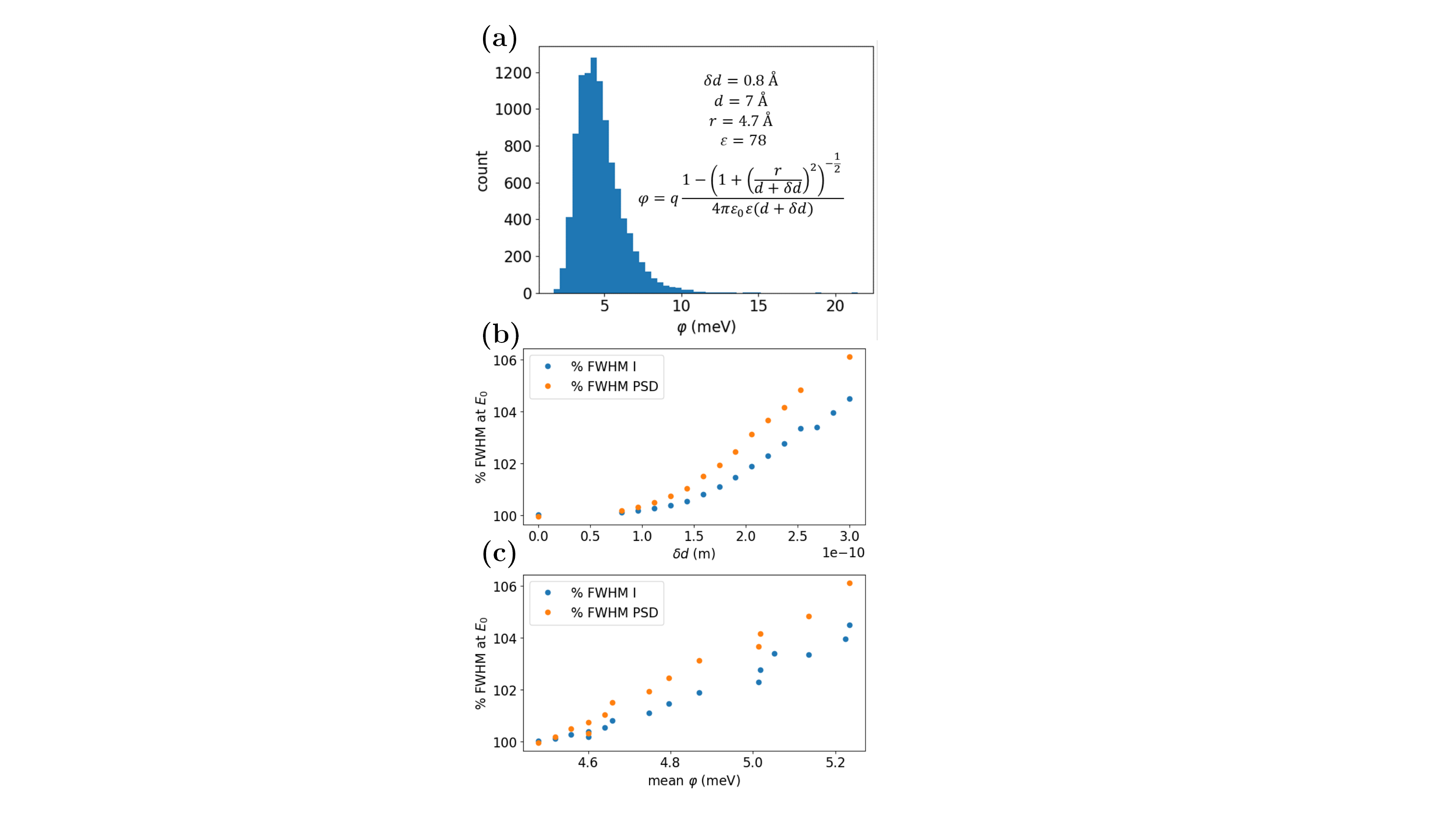}%
\caption{(a) Distribution of $\varphi$ corresponding to a normal distribution of $d$ ($\delta d = 7 \AA$ shown here) in water. Variation of the FWHM of CV and voltnoisogram versus (b) $\delta d$ and the corresponding mean value of $\varphi$ (c).\label{phi_H2O}}
\end{figure}

\begin{figure}[h]
\includegraphics[trim={4cm 0 5cm 0},clip,width=\textwidth]{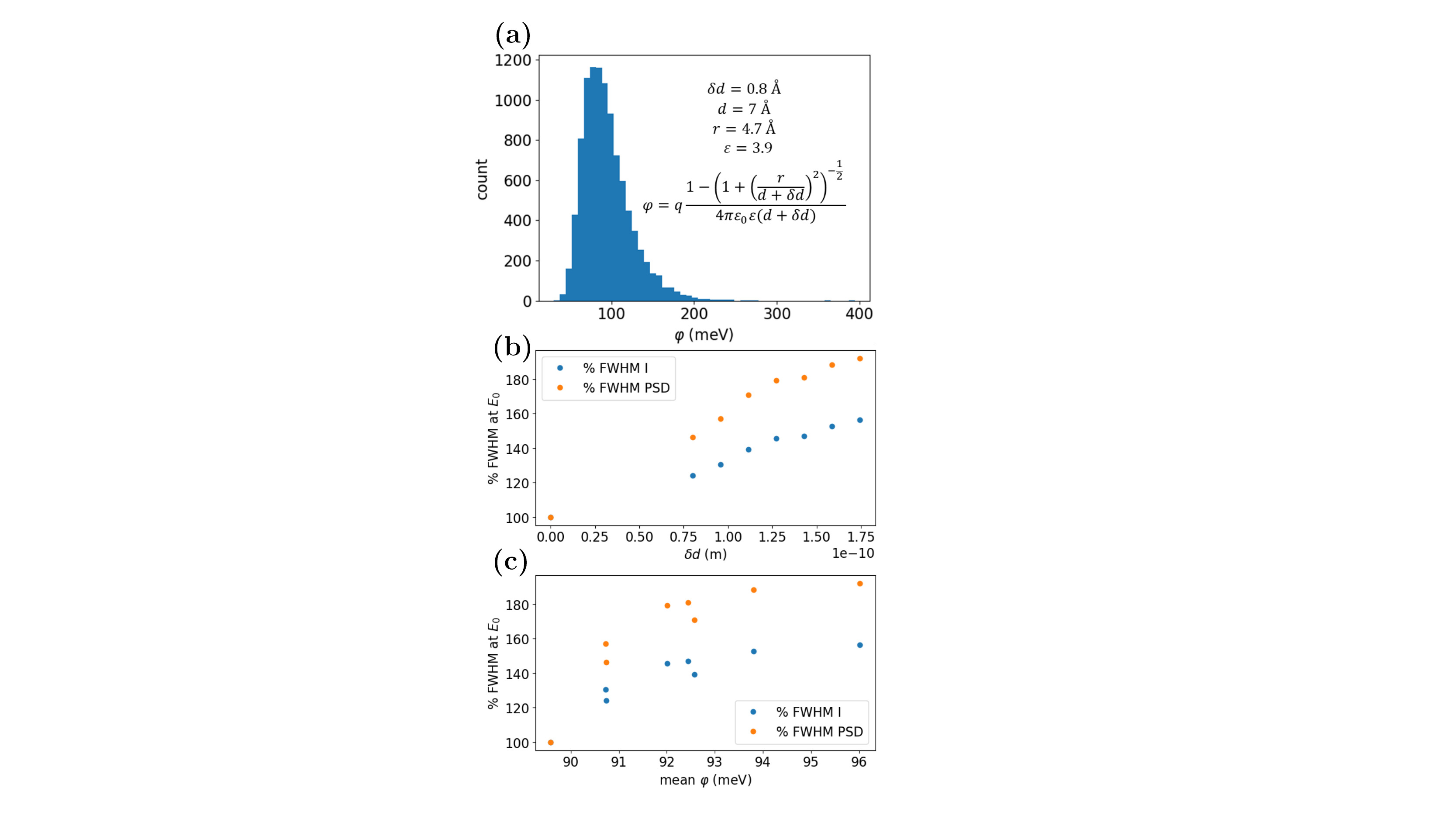}%
\caption{(a) Distribution of $\varphi$ corresponding to a normal distribution of $d$ ($\delta d = 7 \AA$ shown here) in silicon dioxide. Variation of the FWHM of CV and voltnoisogram versus (b) $\delta d$ and the corresponding mean value of $\varphi$ (c).\label{phi_SiO2}}
\end{figure}

Looking at the impact of the permittivity of water $\varepsilon_{H_2O}$ through the electrostatic interactions $\varphi$ such as defined by Eq. \ref{phi} allows to see its effect on current CV and PSD voltnoisograms. To this end, we carried out simulations with normal distributions of the intermolecular distance $d$ with different standard deviations $\delta d$, such as done in \cite{trasobares_estimation_2017}. The impact on current CV and noise voltnoisograms is estimated with their full width at half maximum (FWHM). One set was simulated with $\varepsilon_{H_2O}$ (Figure \ref{phi_H2O}), the other with $\varepsilon_{SiO_2}$ (Figure \ref{phi_SiO2}), showing variations of few percent and tens of percent respectively in the FWHM for both current and PSD.

\clearpage
\section{Discussion on the denomination of the noise}
We acknowledge that the denomination of the noise in the present study may be open for debate. The shot noise is most commonly measured in two-electrodes devices leading to the equation $S = 2qI$, with $I$ the current flowing between the two electrodes and here $I=0$ as a result of two currents canceling each other in this single electrode configuration. However, the original nature of the shot noise lies in the quantization of the electric charge and was previously understood as is by the electrochemists' community \cite{schottky_uber_1918,singh_stochastic_2016}, which is what motivates our proposal of ``Electrochemical shot noise'' for the noise in the present study.

\clearpage
\bibliography{Paper_noise_electrochemistry}